\documentclass[%
 aip,
 amsmath,amssymb,
 reprint,%
]{revtex4-1}

\usepackage{graphicx}
\usepackage{dcolumn}
\usepackage{bm}

\usepackage[utf8]{inputenc}
\usepackage[T1]{fontenc}
\usepackage{mathptmx}
\usepackage{etoolbox}

\makeatletter
\def\@email#1#2{%
 \endgroup
 \patchcmd{\titleblock@produce}
  {\frontmatter@RRAPformat}
  {\frontmatter@RRAPformat{\produce@RRAP{*#1\href{mailto:#2}{#2}}}\frontmatter@RRAPformat}
  {}{}
}%
\makeatother
\begin{document}

\preprint{AIP/123-QED}

\title{Computational modeling of the physical features that influence breast cancer invasion into adipose tissue}
\author{Yitong Zheng}
\affiliation{Department of Mechanical Engineering \& Materials Science, Yale University, New Haven, Connecticut 06520}
\affiliation{Integrated Graduate Program in Physical and Engineering Biology, Yale University, New Haven, Connecticut 06520}

\author{Dong Wang}
\affiliation{Department of Mechanical Engineering \& Materials Science, Yale University, New Haven, Connecticut 06520}

\author{Garrett Beeghly}
\affiliation{Nancy E. and Peter C. Meinig School of Biomedical Engineering, Cornell University, Ithaca, New York 14853}

\author{Claudia Fischbach}
\affiliation{Nancy E. and Peter C. Meinig School of Biomedical Engineering, Cornell University, Ithaca, New York 14853}

\author{Mark D. Shattuck}
\affiliation{Benjamin Levich Institute and Physics Department, City College of New York, New York, New York 10031}

\author{Corey S. O'Hern}
\affiliation{Department of Mechanical Engineering \& Materials Science, Yale University, New Haven, Connecticut 06520}
\affiliation{Integrated Graduate Program in Physical and Engineering Biology, Yale University, New Haven, Connecticut 06520}
\affiliation{Department of Physics, Yale University, New Haven, Connecticut 06520}

\date{\today}

\begin{abstract}
Breast cancer invasion into adipose tissue strongly influences disease progression and metastasis. The degree of cancer cell invasion into adipose tissue depends on numerous biochemical and physical properties of cancer cells, adipocytes, and other key components of adipose tissue. We model breast cancer invasion into adipose tissue as a physical process by carrying out simulations of active, cohesive spherical particles (cancer cells) invading into confluent packings of deformable polyhedra (adipocytes). We quantify the degree of invasion by calculating the interfacial area $A_t$ between cancer cells and adipocytes. We determine the long-time value of $A_t$ versus the activity and strength of the cohesion between cancer cells, as well as mechanical properties of the adipocytes and extracellular matrix (ECM) in which the adipocytes are embedded. We show that the degree of invasion collapses onto a master curve by plotting it versus a dimensionless energy scale $E_c$, which grows linearly with mean-square fluctuations and persistence time of the cancer cell velocities, is inversely proportional to the pressure of the system, and has an offset that increases with the cancer cell cohesive energy.  The condition, $E_c \gg 1$, indicates that cancer cells will invade the adipose tissue, whereas for $E_c \ll 1$, the cancer cells and adipocytes remain demixed.  We also show that constraints on adipocyte positions by the ECM decrease $A_t$ relative to that obtained for unconstrained adipocytes.  Finally, spatial heterogeneity in structural and mechanical properties of the adipocytes in the presence of ECM impedes invasion relative to adipose tissue with uniform properties.
\end{abstract}

\maketitle

\section*{Introduction}
Breast cancer cell invasion into mammary adipose tissue is a key step toward disease progression and metastasis.\cite{majidpoor2021steps,novikov2021mutational} Gaining insight into the biochemical and biophysical processes by which cancer cells invade adipose tissue is crucial for advancing our knowledge of breast cancer and improving treatments. Indeed, the dynamics of breast cancer cell invasion are known to depend on important biochemical factors in the tumor microenvironment (TME).\cite{milotti2017pulsation,marti2005angiogenesis,Sauer2023} For example, cancer cells secrete chemical signals that promote the growth of blood vessels, which in turn provide oxygen and other nutrients to the tumor.\cite{marti2005angiogenesis} Oscillations in the oxygen concentration in the TME caused by collective diffusion affect tumor growth and invasion.\cite{milotti2017pulsation} Moreover, genetic and chemical cues can trigger the epithelial-mesenchymal transition (EMT), which decreases the cohesion between cancer cells and increases their motility.\cite{Sauer2023} In the context of adipose tissue, interactions between cancer cells and adipocytes cause lipid loss and de-differentiation in adipocytes, which promote invasion by altering cancer cell metabolism and reconfiguring the stroma.\cite{Zhu2022,zhang2018adipocyte}

However, the {\it physical} properties of the cancer cells and adipose tissue also play important roles in mediating breast cancer invasion.\cite{Friedl2012,Janiszewska2020,li2017macrophage,Krakhmal2015,Gon2021,Zhu2022,seo2015obesity} In adipose tissue, ECM fiber alignment is influenced by adipocyte properties and regulates  tumor cell invasion by controlling migration persistence.\cite{seo2015obesity,shimpi2023phosphoproteomic} More persistently moving cancer cells, in turn, can invade adipose tissue more rapidly.  Cancer cell cohesion also influences tumor growth and invasion.\cite{Friedl2012,Janiszewska2020} Highly cohesive cancer cells do not disconnect from each other and thus do not as easily invade out of the primary tumor. With decreasing cohesion, cancer cells can more readily detach from the tumor and invade distant tissues.\cite{Krakhmal2015,Friedl2012} During invasion, cancer cells navigate between adipocytes which form dense polyhedral packings. Thus, the packing fraction and shape deformation of adipocytes, as well as the porous ECM network between adipocytes also physically influence breast cancer invasion.\cite{wolf2013physical,seo2015obesity} 

\begin{figure*}[!htbp]
\centering
  \includegraphics[width=17cm]{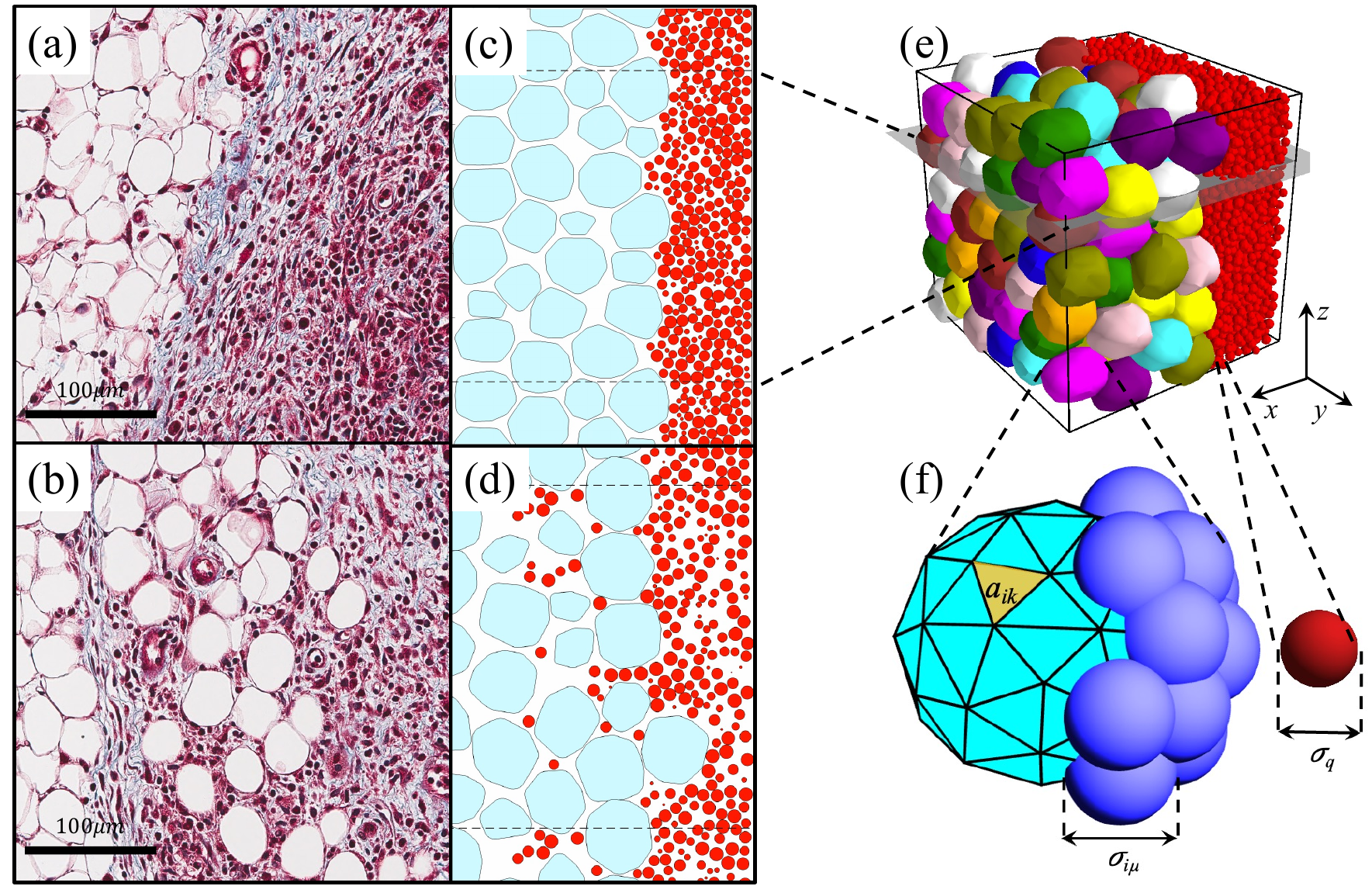}
\caption{(a) and (b) Histological slices of the interface between a mammary tumor and adipose tissue in mice. Cancer cell nuclei are colored reddish-purple and the large white regions are adipocytes. (a) Noninvaded adipose tissue, characterized by a  smooth interface between the tumor and adipose tissue. (b) Invaded adipose tissue with a rough interface. (c) and (d) 2D slices in the $xy$-plane from 3D DPM simulations in (e), both (c) before and (d) after invasion. Red circles represent cancer cells and adipocytes are shown as smooth, cyan-colored deformable polygons. Dashed lines indicate periodic boundaries in the $y$-direction. (e) Initial configuration of de-mixed cancer cells (red spheres) and adipocytes (smooth, multi-colored deformable polyhedra) in simulations with $N_a=128$ and $N_c=7000$. (f) Adipocytes are modeled as deformable polyhedra with vertices of diameter $\sigma_{i\mu}$ and a triangular mesh with  individual triangle area $a_{ik}$. Cancer cells are modeled as soft spheres with diameter $\sigma_q$.}
  \label{fig:model}
\end{figure*}

In this work, we investigate the key physical variables that influence cancer cell invasion into adipose tissue. We leverage the recently developed deformable particle model (DPM)\cite{ton2024mechanical,Wang2021,Cheng2022,Boromand2018,Treado2022,Treado2021,manning2023essay} and carry out DPM simulations of self-propelled, cohesive spherical particles (cancer cells) invading adipose tissue, which we model as dense packings of deformable polyhedra that are tethered to each other to mimic adhesion to the ECM. The systems are de-mixed when initialized, with a smooth interface between the cancer cells and adipose tissue. During simulations, we quantify the invasion process by calculating the interfacial area shared by the cancer cells and adipocytes as a function of time. In our computational studies, we do not consider pressure gradients caused by cell proliferation but instead focus on invasion due to cancer cell migration. These novel simulations allow us to study the degree of tumor invasion as a function of the cohesion and motility of the cancer cells, as well as the structural and mechanical properties of the adipocytes and ECM. 

As shown in previous studies, we find that strong cohesion between cancer cells hinders tumor invasion, while increased cell motility promotes invasion. We identify a characteristic dimensionless energy scale $E_c$ that controls the degree of invasion. $E_c$ increases with the speed and persistence of cancer cell migration and decreases with the pressure in the system. To model the adhesion of adipocytes to the ECM, we add springs connecting neighboring adipocytes to constrain their motion. We show that tethering the adipocytes decreases invasion compared to untethered adipocytes by an amount that is controlled by the packing fraction of the adipose tissue. Finally, we show that spatial heterogeneity in the mechanical properties of adipose tissue impedes invasion relative to adipose tissue with uniform mechanical properties.\par

We describe these findings in more detail in the remaining three sections of the article. In the Methods section, we outline the DPM shape-energy function for adipocytes and the soft particle model for the cancer cells, the interactions between adipocytes, between cancer cells, and between adipocytes and cancer cells, as well as the integration of the equations of motion for the cancer cells and adipocytes.  In the Results section, we define the interfacial area $A_t$ shared by the cancer cells and adipocytes and show how it depends on the dimensionless energy scale $E_c$. We then show that $A_t$ decreases with tethering and heterogeneity in the mechanical properties of the adipocytes. In the Conclusions and Discussion section, we emphasize that the physical properties of adipose tissue, not only cancer cells, influence breast cancer invasion. We also propose future experimental studies of breast cancer cell invasion into soft spherical particles embedded in collagen fibers that will allow us to determine values of $E_c$ for systems that mimic the physical properties of adipose tissue. We also include three Appendices. In Appendix A, we show that our chosen reference shape parameter $\mathcal{A}_0$ for adipocytes is similar to those found in recent experimental studies of adipose tissue. In Appendix B, we further describe the interactions between cancer cells and adipocytes in simulations. Finally, in Appendix C, we describe how the interfacial area $A_t$ between cancer cells and adipocytes is normalized to remove the effects of dilation at large $E_c$. 

\begin{figure}[!htbp]
\centering
  \includegraphics[width=8cm]{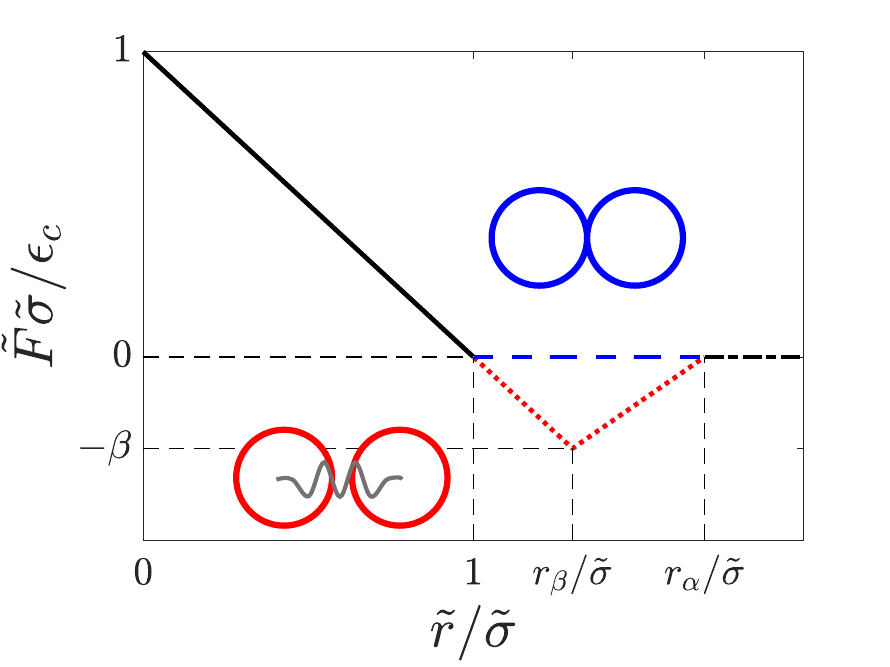}
\caption{The pairwise interaction force $\tilde{F}$ plotted as a function of the separation $\tilde{r}$ between different cell types. For adipocyte-adipocyte interactions, $\tilde{F}=F_{ij\mu\nu}$, $\tilde{r}=r_{ij\mu\nu}$, and $\tilde{\sigma}=\sigma_{ij\mu\nu}$, where $i,j=1,\ldots,N_a$ are the adipocyte indices and $\mu,\nu=1,\ldots,N_v$ are the vertex indices. $N_a$ and $N_v$ are the number of adipocytes in the simulation box and the number of vertices on an adipocyte, respectively. For adipocyte-cancer cell interactions, $\tilde{F}=F_{i\mu q}$, $\tilde{r}=r_{i\mu q}$, and $\tilde{\sigma}=\sigma_{i\mu q}$, where $q=1,\ldots,N_c$. $N_c$ is the number of cancer cells in the simulation box.  Both adipocytes and cancer cells interact via repulsive linear spring forces when they overlap (black solid line) and do not interact when they are out of contact (blue dashed and black dash-dotted lines). For the cancer cell-cancer cell interactions, $\tilde{F}=F_{qs}$, $\tilde{r}=r_{qs}$, and $\tilde{\sigma}=\sigma_{qs}$. These interactions include linear repulsion when two cancer cells overlap (black solid line), short-range attraction for $1 < {\tilde r}/{\tilde \sigma} < r_{\alpha}/{\tilde \sigma}$  (red dotted line), and no interactions for ${\tilde r}/{\tilde \sigma} > r_{\alpha}/{\tilde \sigma}$ (black dash-dotted line).}
  \label{fig:interaction}
\end{figure}
\begin{figure*}[!htbp]
\centering
  \includegraphics[width=15cm]{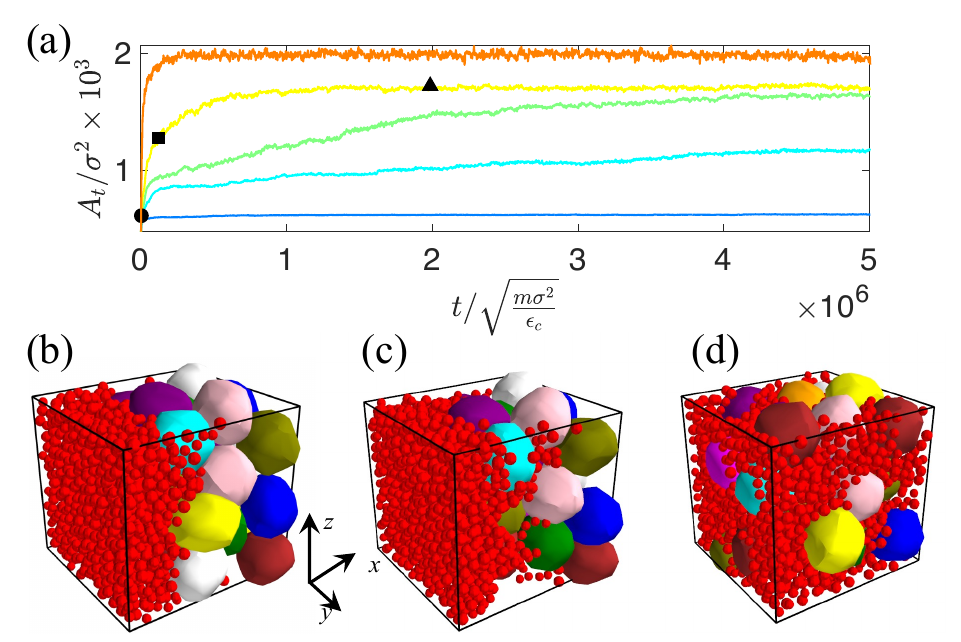}
\caption{(a) Interfacial surface area $A_t$ (in units of $\sigma^2$) shared between the cancer cells and adipocytes plotted as a function of time (in units of $\sqrt{m \sigma^2/\epsilon_c}$). We show $A_t$ for five values of $\frac{k_bT}{P\sigma^3}=1.25 \times 10^{-3}$, $7.89 \times 10^{-2}$, $1.25 \times 10^{-1}$, $1.98 \times 10^{-1}$, and $1.25$ increasing from blue to red. Simulation snapshots for different $A_t$ values and times from (a) are shown in (b)-(d): (b) $A_t/\sigma^2=610$, (c) $1280$, and (d) $1720$. The time of each snapshot is indicated by the corresponding shapes shown in (a): (b) triangle, (c) square, and (d) circle.}
  \label{fig:invasion}
\end{figure*}

\section*{Methods}
In this section, we describe the methods for simulating breast cancer cell invasion into adipose tissue. First, we introduce the shape-energy function for the 3D deformable particle model, which is used to model dense packings of adipocytes. We then describe the self-propelled, attractive soft particle model for the cancer cells and the repulsive contact interactions between cancer cells and adipocytes. We outline the equations of motion for the cancer cells and adipocytes and describe the methods we use to integrate them to calculate cell trajectories. Finally, we specify the initial configuration as a de-mixed system with a smooth, planar interface between cancer cells and adipocytes, which also includes physical walls in one direction to control  pressure, and periodic boundary conditions in the other two directions. 

\subsection*{Deformable particle model for adipocytes}
To model dense packings of adipocytes, which form faceted cell-cell contacts under compression, we leverage the recently developed 3D deformable particle model. Specifically, we consider adipocytes as deformable polyhedra composed of $N_v=42$ vertices belonging to a mesh with $N_f=80$ triangular faces, which is sufficient to describe the shape fluctuations observed in adipocytes. (See Fig. \ref{fig:model} (f).) The shape energy for the \textit{i}th adipocyte as a function of the positions of its vertices $\mathbf{r}_{i1}, \ldots, \mathbf{r}_{iN_v}$ is:
\begin{equation}\label{Udpm}
U_i= \frac{\epsilon_v}{2} \left(1 - \frac{v_i}{v_{i0}}\right)^2 + \frac{\epsilon_a}{2} \sum_{k=1}^{N_f}   \left(1 - \frac{a_{ik}}{a_{ik0}}\right)^2.
\end{equation}
The first term imposes a harmonic energy penalty for changes in cell volume $v_i$ from its preferred value $v_{i0}$ and $\epsilon_v$ controls the fluctuations in adipocyte volume. We set $\epsilon_v$ to be much larger than the other energy scales in the system so that the adipocytes cannot change their volume. The second term imposes a harmonic energy penalty for deviations in the area $a_{ik}$ of each triangular face from its preferred value $a_{ik0}$ and $\epsilon_a$ controls the fluctuations in the adipocyte surface area. In general, 3D deformable particles without bending energy are floppy and can change shape without increasing energy. (In particular, deformable particles with $N_v=42$ vertices and $N_f=80$ faces have $39$ zero-energy modes of the dynamical matrix for the shape-energy function in Eq.~\ref{Udpm}.\cite{Wang2021}) We quantify adipocyte shape using the dimensionless shape parameter $\mathcal{A} = \frac{\left(\sum_{k=1}^{N_f} a_{ik}\right)^{\frac{3}{2}}}{6\sqrt{\pi}v_i}$. The shape parameter for a sphere is $1$ and ${\mathcal A} > 1$ for all nonspherical shapes. We set the reference shape parameter $\mathcal{A}_0 = 1.1$ for adipocytes based on experimental observations shown in Appendix A.

\subsection*{Interactions between adipocytes, between cancer cells, and between adipocytes and cancer cells}
The interactions between adipocytes are controlled by the pair forces between their vertices. In particular, the repulsive pair force between vertex $\mu$ on adipocyte $i$ and vertex $\nu$ on adipocyte $j$ is purely repulsive: 
\begin{equation}\label{F1}
{\textbf{F}}_{ij\mu\nu} = 
\begin{cases}
\frac{\epsilon_c}{\sigma_{ij\mu\nu}}\left(1-\frac{r_{ij\mu\nu}}{\sigma_{ij\mu\nu}}\right) {\hat r}_{ij\mu\nu}, & r_{ij\mu\nu}\leq \sigma_{ij\mu\nu},\\
0, & r_{ij\mu\nu}>\sigma_{ij\mu\nu}
\end{cases}
\end{equation}
where $\epsilon_c$ controls the strength of the repulsive forces, ${\hat r}_{ij\mu\nu}={\bf r}_{ij\mu\nu}/r_{ij\mu\nu}$, ${\bf r}_{ij\mu\nu}=\textbf{r}_{i\mu}-\textbf{r}_{j\nu}$ is the separation vector, $r_{ij\mu\nu}=|{\bf r}_{ij\mu\nu}|$ is the distance between vertices, and $\sigma_{ij\mu\nu}=\frac{1}{2}(\sigma_{i\mu}+\sigma_{j\nu})$ is their average diameter. (See Fig.~\ref{fig:distance} in Appendix B.) $\sigma_{i\mu}$ is approximately two times the average cancer cell diameter, which allows us to cover the surface of the adipocytes with vertices to prevent interpenetration by cancer cells and other adipocyte vertices.

The diameter of adipocytes in human breast adipose tissue ranges from $50$ to $150 \mu m$,\cite{parlee2014quantifying} while cancer cells are much smaller, ranging between $10$ and $20 \mu m$.\cite{Truongvo2017}. Since the magnitude of cancer cell shape fluctuations is much smaller than that for adipocytes, we do not consider shape fluctuations of cancer cells and instead model them as soft spheres. We assume that the force on vertex $\mu$ on adipocyte $i$ from cancer cell $q$ is purely repulsive: 
\begin{equation}\label{F3}
{\bf F}_{i\mu q} = 
\begin{cases}
\frac{\epsilon_c}{\sigma_{i\mu q}}\left(1-\frac{r_{i\mu q}}{\sigma_{i\mu q}}\right){\hat r}_{iuq}, & r_{i\mu q}\leq \sigma_{i\mu q},\\
0, & r_{i\mu q}>\sigma_{i\mu q}
\end{cases}
\end{equation}
where ${\hat r}_{i\mu q}={\bf r}_{i\mu q}/r_{i\mu q}$, ${\bf r}_{i\mu q}={\bf r}_{i\mu}-{\bf r}_{q}$ is the separation vector, $r_{i\mu q}=|{\bf r}_{i\mu q}|$ is the distance between vertex $\mu$ on adipocyte $i$ and cancer cell $q$, $\sigma_{i\mu q}=\frac{1}{2}(\sigma_{i\mu}+\sigma_{q})$, and $\sigma_q$ is the diameter of the $q$th cancer cell.

For the interactions between cancer cells, we include contact repulsions as well as short-range attractions. We assume that the pair force on cancer cell $q$ from cancer cell $s$ obeys:
\begin{equation}\label{F2}
{\bf F}_{qs} = 
\begin{cases}
\frac{\epsilon_c}{\sigma_{qs}}(1-\frac{r_{qs}}{\sigma_{qs}}){\hat r}_{qs}, & r_{qs}\leq r_\beta\\
\frac{\epsilon_c}{\sigma_{qs}}\frac{r_\beta-\sigma_{qs}}{r_\alpha-r_\beta}\frac{r_{qs}-r_\alpha}{\sigma_{qs}} {\hat r}_{qs}, & r_\beta<r_{qs}\leq r_\alpha, \\
0, & r_{qs}> r_\alpha
\end{cases}
\end{equation}
where ${\hat r}_{qs} = {\bf r}_{qs}/r_{qs}$, ${\bf r}_{qs}={\bf r}_q -{\bf r}_s$ is the separation vector, $r_{qs}=|{\bf r}_{qs}|$ is the distance between cancer cells $q$ and $s$, and $\sigma_{qs}=\frac{1}{2}(\sigma_q+\sigma_s)$ is the average diameter of the cancer cells. As shown in Fig.~\ref{fig:interaction}, the pair force between cancer cells is repulsive for $r_{qs} < \sigma_{qs}$, and attractive for $1 < r_{qs}/\sigma_{qs} < r_{\alpha}/\sigma_{qs}$ with maximum attractive force $F_{qs}=\epsilon_c \beta/\sigma_{qs}$ at $r_{qs}=r_{\beta}$, where $r_{\alpha}/\sigma_{qs} = 1+\alpha$ and $r_{\beta}/\sigma_{qs} = 1+\beta$. We vary the depth of the attractive potential $\beta$ from $10^{-4}$ to $10^{-2}$ and set the range of attractive interactions at $\alpha=0.2$. To prevent partial crystallization of the cancer cells during invasion, we assume that the cancer cells are bidisperse in size where half of the cells have a smaller diameter $0.9 \sigma$ and the other half have a larger diameter $1.1 \sigma$, where $\sigma$ is the average diameter of the cancer cells. The ratio of the adipocyte diameter (including the vertex radii) and the average cancer cell diameter $\sigma$ is approximately $6$, which matches recent experimental studies.\cite{parlee2014quantifying,Truongvo2017}

\subsection*{Equations of motion for cancer cells and adipocytes}

To determine the spatiotemporal trajectories of adipocytes and cancer cells, we integrate Newtwon's equations of motion. For the cancer cells, we consider both active and passive models. For the active model, the equation of motion for the position ${\bf r}_q$ of cancer cell $q$ is
\begin{equation}\label{EoM}
\begin{aligned}
m\frac{d^2 \textbf{r}_q}{dt^2} = &\sum^{N_c}_{s\ne q=1} {\bf F}_{qs} - \sum_{i=1}^{N_a}\sum^{N_v}_{\mu=1} {\bf F}_{i\mu q} \\&-\sum_{w=l}^r \textbf{F}_{wq}-\gamma \textbf{v}_q + f_0\hat{n}_q,
\end{aligned}
\end{equation}
where $m$ is the mass of the cancer cells, ${\bf F}_{i\mu q}$ is the pair force between cancer cells and the vertices of the adipocytes (Eq.~\ref{F3}), and ${\bf F}_{qs}$ is the pair force between cancer cells (Eq.~\ref{F2}). ${\bf F}_{wq}$ indicates the repulsive forces between physical walls and the $q$th cancer cell and is defined in Eq.~\ref{wq}. The equation of motion for active cancer cells also includes a damping force from the surrounding environment that is proportional to the cancer cell velocity ${\bf v}_q$ with damping coefficient $\gamma \sigma/\sqrt{m \epsilon_c}=0.2$ and a self-propulsion term with active force magnitude $f_0$. The direction of the active force ${\hat n}_q$ obeys rotational diffusion\cite{Debets2021,shaebani2020computational}, 
\begin{equation}\label{ni}
    \frac{d\hat{n}_q}{d t} = {\bf \chi}_q \times \hat{n}_q,
\end{equation}
where ${\bf \chi}_q$ is a vector with $x$-, $y$-, and $z$-components that are Gaussian random numbers with zero mean and satisfy
\begin{equation}\label{chi}
\left<{\bf \chi}_q(t'){\bf \chi}_s(t'+t)\right>_{t'}=\tau^{-1}_p \textbf{I}\delta_{qs}\delta(t),    
\end{equation}
where $\left< .\right>_{t'}$ is an average over time origins, $\textbf{I}$ is the identity matrix, $\delta_{qs} = 1$ when $q=s$ and $0$ otherwise, and $\delta(.)$ is the Dirac $\delta$-function. \( \tau_p \) is the persistence time of the active force as defined through the autocorrelation function of \( \hat{n}_q \): 
\begin{equation}\label{aveN}
\left< \hat{n}_q(t') \cdot \hat{n}_q(t' + t) \right>_{q,t'} = \exp\left(-\frac{t}{\tau_p}\right),
\end{equation}
where $\left< .\right>_{q,t'}$ is an average over cancer cells and time origins. Persistence in cancer cell migration originates from the fact that the migration direction and speed are strongly correlated with the local alignment of ECM fibers~\cite{Wang2018}. In this model, we include a background temperature that captures the random motion of the cancer cells and persistence time $\tau_p$ that captures the correlation between cancer cell velocity and local ECM fiber alignment.
 
We also consider a passive model for cancer cell migration to better understand the role of persistent cancer cell motion on breast cancer invasion. For the passive model, we employ the Langevin equation of motion for the $q$th cancer cell:
\begin{equation}\label{Langevin}
\begin{aligned}
m\frac{d^2 \textbf{r}_q}{d t^2} =& \sum_{s\neq q=1}^{N_c} {\bf F}_{qs} - \sum_{i=1}^{N_a}\sum_{\mu=1}^{N_v} {\bf F}_{i\mu q}\\& -\sum_{w=l}^r \textbf{F}_{wq}-\gamma \textbf{v}_q + \sqrt{2k_bT_0\gamma m} {\bf \eta}_q(t),
\end{aligned}
\end{equation}
where the first and second terms include interactions among cancer cells and between cancer cells and adipocytes. As in Eq.~\ref{EoM}, the equation of motion for passive cancer cells includes a damping force proportional to the velocity ${\bf v}_q$ of cancer cell $q$ with damping coefficient $\gamma$. The last term is the noise term with strength $\sqrt{2 k_b T_0 \gamma m}$ that sets the target temperature $T_0$. ${\bf \eta}(t)$ is a vector with $x$-, $y$-, and $z$-components that are Gaussian random numbers with zero mean and unit variance.

We employ damped equations of motion for the adipocytes, i.e. the position of vertex $\mu$ on the $i$th adipocyte obeys: 
\begin{equation}\label{EoM2}
\begin{aligned}
m\frac{d^2 \textbf{r}_{i\mu}}{d t^2} = & \sum_{j\neq i=1}^{N_a} \sum_{\nu=1}^{N_v} {\bf F}_{ij\mu\nu} + \sum_{q=1}^{N_c} {\bf F}_{i\mu q} \\& -\sum_{w=l}^r \textbf{F}_{wi\mu}-\gamma \textbf{v}_{i\mu} -\nabla_{\textbf{r}_{i\mu}}U_{i},
\end{aligned}
\end{equation}
where $m$ is the mass and $\gamma$ is the damping constant associated with each adipocyte vertex. The first and second terms give the repulsive pair forces between vertices $\mu$ and $\nu$ on different adipocytes $i$ and $j$ and between vertex $\mu$ on adipocyte $i$ and cancer cell $q$. ${\bf F}_{wi\mu}$ is the repulsive force between the physical walls and vertex $\mu$ on the $i$th adipocyte, which is defined in Eq. \ref{wi}. The fifth term gives the mechanical forces generated from the shape-energy function for the $i$th adipocyte in Eq. (\ref{Udpm}).

We integrate the equations of motion for the adipocyte vertices and cancer cells (i.e. Eqs.~\ref{EoM}, \ref{Langevin}, and~\ref{EoM2}), using the modified velocity Verlet integration scheme with time step $\Delta t=5\times10^{-2}\sqrt{\frac{m\sigma^2}{\epsilon_c}}$.

\subsection*{Initialization and boundary conditions for invasion simulations}

We initialize the cancer cells and adipocytes in a dilute, de-mixed state. The simulation box is cubic with initial side length $L_0=70\sigma$, $N_a=28$ adipocytes, $N_c=1500$ cancer cells, and initial total packing fraction $\phi=0.01$.  (Note the total packing fraction can be written in terms of the packing fractions for cancer cells and adipocytes, separately, $\phi=\phi_c +\phi_a$.) The cancer cells and adipocytes are placed randomly in the simulation box such that the $x$-components of the cancer cell centers of mass satisfy $2 \sigma < x_q < 12\sigma$ and the $x$-components of the adipocyte centers of mass satisfy $16 \sigma < X_i < 68\sigma$. We then perform athermal, isotropic compression, applying successive compression steps $\Delta \phi/\phi=0.03$. Compression is followed by energy minimization until the final packing fraction of the system reaches $\phi=(\frac{\pi}{6}\sum_{q=1}^{N_c}\sigma_q^3+\sum_{i=1}^{N_a} V_i)/L_f^3=0.72$, where $L_f=34\sigma$ is the final side length of the box, and $V_i$ is the total volume of the $i$th adipocyte, i.e. $v_i$ plus the additional volumes from the vertices that do not overlap the polyhedron. (See Fig. \ref{fig:model} (e).) 

During compression, we apply periodic boundary conditions in the $y$- and $z$-directions. In the $x$-direction, we include two physical walls, a ``right" wall at $x_r=L_f$ and a ``left" wall at $x_l=0$. The right wall is stationary, while the left wall moves in the $x$-direction to maintain fixed pressure $P$. The equation of the motion for the left wall position is:
\begin{equation}
    m\frac{d^2x_l}{dt^2}=\sum_{q=1}^{N_c}\textbf{F}_{lq}\cdot\hat{x}+\sum_{i=1}^{N_a}\sum_{\mu=1}^{N_v}\textbf{F}_{li\mu}\cdot\hat{x}+PL_f^2-\gamma \textbf{v}_l,
\end{equation}
where $m$ is the mass and $\gamma$ is the damping coefficient for the left wall, $v_l$ is the speed of the left wall. The forces that act on the wall from the $q$th cancer cell and $\mu$th vertex on the $i$th adipocyte are:
\begin{equation}\label{wq}
{\textbf{F}}_{wq} = 
\begin{cases}
\frac{\epsilon_c}{\sigma_q}(1-\frac{x_{wq}}{\sigma_{q}})\hat{x}, & x_{wq}\leq \sigma_{q},\\
0, & x_{wq}>\sigma_{q},
\end{cases}
\end{equation}
and
\begin{equation}\label{wi}
{\textbf{F}}_{wi\mu} = 
\begin{cases}
\frac{\epsilon_c}{\sigma_{i\mu}}(1-\frac{x_{wi\mu}}{\sigma_{i\mu}})\hat{x}, & x_{wi\mu}\leq \sigma_{i\mu},\\
0, & x_{wi\mu}>\sigma_{i\mu},
\end{cases}
\end{equation}
where $x_{wq}=x_{w}-x_q$ and $x_{wi\mu}=x_{w}-x_{i\mu}$. (Note that the subscript $w=l$,$r$ indicates the left and right walls, respectively.) We quantify the temperature of the system using the kinetic energy of the cancer cells:
\begin{equation}
    k_bT = \frac{2m}{3N_c}\sum_{q=1}^{N_c}v_q^2.
\end{equation}
For the passive model of cancer cells, the temperature of the cancer cells $T_c$ and adipocytes $T_a$ equilibrate. For the active model of cancer cells, the temperature of the cancer cells is larger than that for the adipocytes. In the systems considered here, $1 < T_c/T_a < 1.2$. 

\begin{figure}[!htbp]
\centering
  \includegraphics[width=8cm]{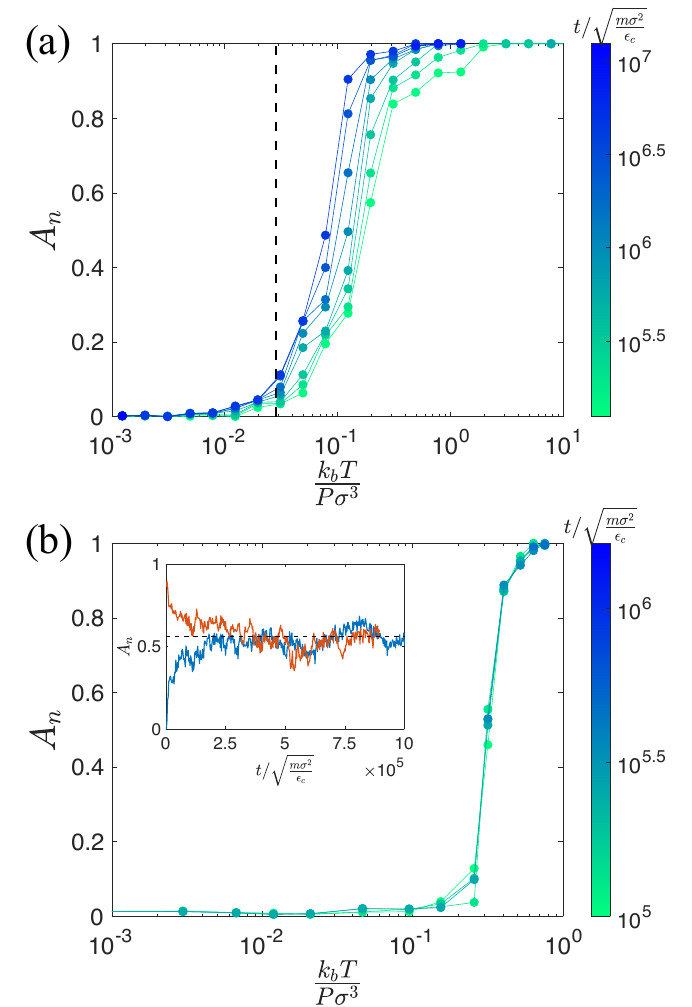}
\caption{Normalized interfacial area $A_{n}$ between the cancer cells and adipocytes (defined in Appendix B) plotted as a function of $k_bT/(P\sigma^3)$ at different times $t$, over a wide range of temperatures $T$, and where pressure $P=1.3 \times 10^{-3} \frac{\epsilon_c}{\sigma^3}$  for (a) purely repulsive, passive cancer cells with $\beta = 0$ and $\tau_p = 0$ and (b) attractive, active cancer cells with $\beta=10^{-2}$ and $\tau_p/\sqrt{\frac{m\sigma^2}{\epsilon_c}} = 25$. In (a), the vertical line indicates the glass transition temperature $k_b T_g/(P \sigma^3) \approx 0.03$ below which the structural relaxation $\tau_r \rightarrow \infty$. In (b), the inset shows that the equilibrium value of $A_n \approx 0.55$ (black dashed line) for $k_bT/(P\sigma^3)\approx0.3$ is independent of the initial state since the same value of $A_n$ is obtained when the system is initialized in a de-mixed state (blue line) and a mixed state (red line). In both (a) and (b), time increases from green to blue.}
  \label{fig:transition}
\end{figure}

\begin{figure}[!htbp]
\centering
  \includegraphics[width=8cm]{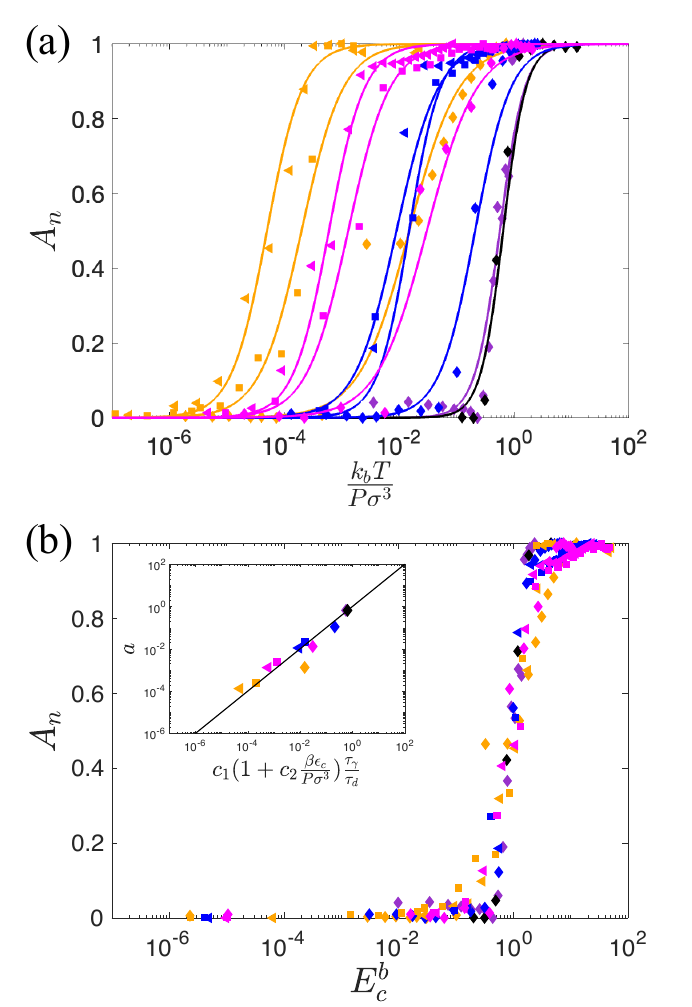}
\caption{(a) Normalized interfacial area $A_n$ is plotted versus $\frac{k_bT}{P\sigma^3}$ for several values of  attraction strength $\beta$ and persistence time $\tau_p$ of the cancer cells. The symbols indicate the attraction strength: $\beta =10^{-4}$ (triangles), $10^{-3}$ (squares), and $10^{-2}$ (diamonds). The colors indicate the persistence time: $\tau_p/\sqrt{\frac{m\sigma^2}{\epsilon_c}}=0.02$ (purple),  $25$ (blue), $250$ (magenta), and $2500$ (orange). We also show data for passive cancer cells with $\tau_p=0$ and $\beta = 10^{-2}$ as filled black diamonds. The solid lines are fits to Eq. \ref{fitA}. (b) $A_n$ is plotted versus the dimensionless energy scale $E_c^b$ for the same data in (a). The inset shows the relationship between the parameter $a$ in Eq.~\ref{fitA} and $c_1\left(1+c_2\frac{\beta\epsilon_c}{P\sigma^3}\right)\frac{\tau_\gamma}{\tau_d}$. The solid black line indicates $a=c_1\left(1+c_2\frac{\beta\epsilon_c}{P\sigma^3}\right)\frac{\tau_\gamma}{\tau_d}$.}
  \label{fig:effects}
\end{figure}

\begin{figure}[!htbp]
\centering
  \includegraphics[width=8cm]{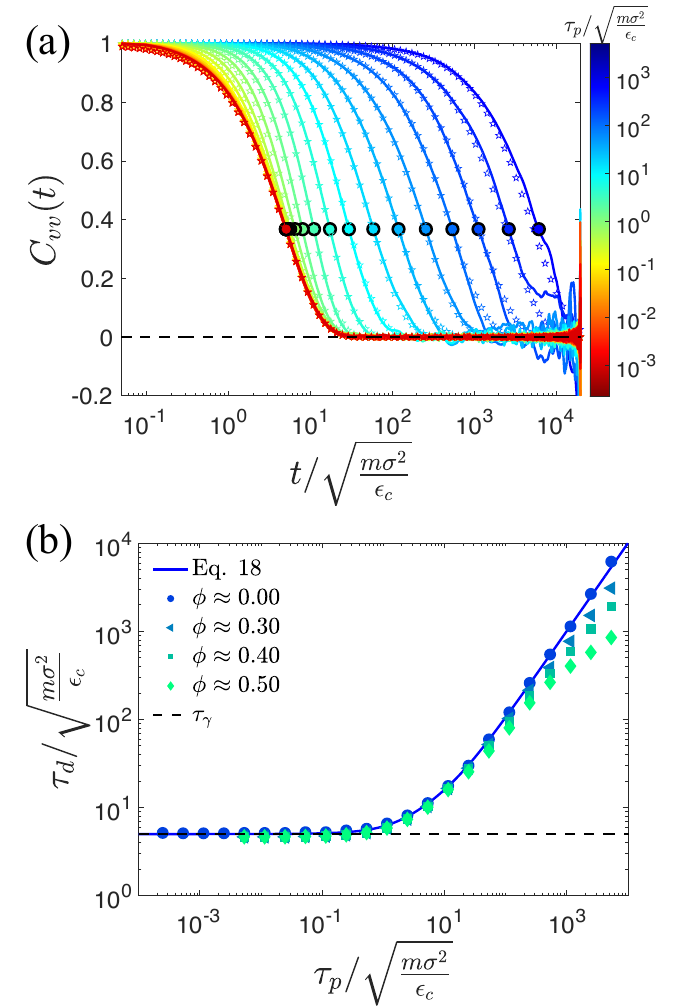}
\caption{(a) The velocity autocorrelation function $C_{vv}(t)$ at different $\tau_p$ and $\phi_c\approx0$. The filled circles indicate the velocity decorrelation times $\tau_d$ at which $C_{vv}(\tau_d)=e^{-1}$. The colors from red to blue indicate short to long $\tau_p$. Open stars represent Eq.~\ref{Cvv0}. (b) The decorrelation time $\tau_d/\sqrt{\frac{m\sigma^2}{\epsilon_c}}$ plotted versus $\tau_p/\sqrt{\frac{m\sigma^2}{\epsilon_c}}$. The filled symbols indicate $\tau_d$ from simulations of repulsive, active cancer cells without adipocytes at $\phi_c=0$ (circles), $0.3$ (triangles), $0.4$ (squares), and $0.5$ (diamonds). The analytical solution for $C_{vv}(\tau_d)=e^{-1}$ when $\phi_c=0$ (Eq.~\ref{Cvv0}) is shown as a blue solid line\cite{Caprini2021}. The horizontal dashed line indicates $\tau_d = \tau_{\gamma}$.}
  \label{fig:Cvv}
\end{figure}

\section*{Results}

In this section, we describe the results from discrete element method simulations of breast cancer invasion into adipose tissue. First, we describe how we quantify the degree of invasion using the interfacial area $A_t$ between cancer cells and adipocytes. We then calculate $A_t$ as a function of time as the cancer cells and adipocytes evolve from a de-mixed to mixed state over a wide range of pressures and temperatures. If the system is above the glass transition temperature, the purely repulsive, passive cancer cells evolve toward a mixed (invaded) state. For attractive cancer cells, the long-time degree of invasion depends on the strength of the cancer cell cohesion. We identify a dimensionless energy scale, where $E_c >1$ indicates the onset of cancer cell invasion. We find that $E_c$ is inversely proportional to the persistence of cancer cell motion and increases with the strength of attraction. Next, we study the effects of the ECM on cancer invasion by adding linear springs between neighboring adipocytes to constrain their motion. We find that constraining adipocytes does not alter the onset of invasion but decreases the value of $A_t$ relative to untethered adipocytes. Finally, we show that heterogeneity in the local structural and mechanical properties of tethered adipocytes restricts cancer cell invasion.

\subsection*{Quantifying the degree of cancer cell invasion}

As discussed in the Methods section, cancer cells and adipocytes are initialized in a de-mixed state. We then  integrate their equations of motion (i.e. Eqs.~\ref{EoM} or~\ref{Langevin} for cancer cells and Eq.~\ref{EoM2} for adipocytes) to determine their spatiotemporal trajectories over a wide range of temperatures and pressures. To illustrate how we quantify the degree of cancer cell invasion, we first focus on the passive model with purely repulsive interactions for cancer cells (Eq.~\ref{Langevin}). We quantify the invasion process by calculating the interfacial area $A_{t}$ between the cancer cells and adipocytes (as well as between cancer cells and the walls) as a function of time. As discussed in Appendix C, $A_t$ is determined by performing a Voronoi tessellation and identifying the faces of the Voronoi polyhedra that are shared by cancer cells and adipocytes (and cancer cells and the walls). In Fig.~\ref{fig:invasion} (a), we show that the time dependence of $A_t$ strongly varies with $k_b T/(P\sigma^3)$. The minimum possible value of $A_t$ is  $A^{\rm min}_t = 470\sigma^2$, which corresponds to the initial, de-mixed value (Fig.~\ref{fig:invasion} (b)), and the maximum possible value of $A_t$ is $A_t^{\rm max}=2000 \sigma^2$, which corresponds to the surface area of the $N_a$ adipocytes (Fig.~\ref{fig:invasion} (d)). Thus, $A_t$ must exist between these two values during the cancer cell invasion process. To enable comparisons of the invasion process for different system parameters, we normalize $A_t$ by its minimum and maximum values, while accounting for dilation of the system at high temperatures, as discussed in Appendix C. In Fig.~\ref{fig:transition} (a), we show the normalized interfacial area $A_n$ as a function of $k_b T/(P\sigma^3)$, where $0< A_n < 1$, for purely repulsive, passive cancer cells for different times following initialization of the de-mixed state. For each time $t$, $A_n$ is sigmoidal when plotted versus $k_b T/(P\sigma^3)$, but shifts to smaller $k_b T/(P\sigma^3)$ as $t$ increases. These results indicate that purely repulsive, passive cancer cells will mix with adipocytes, reaching $A_n =1$ for all $k_b T/(P\sigma^3)$ that are above the glass transition temperature, where structural relaxation times diverge.

\subsection*{Effects of cancer cell cohesion and persistence on invasion}

\begin{figure}[!htbp]
\centering
\includegraphics[width=8cm]{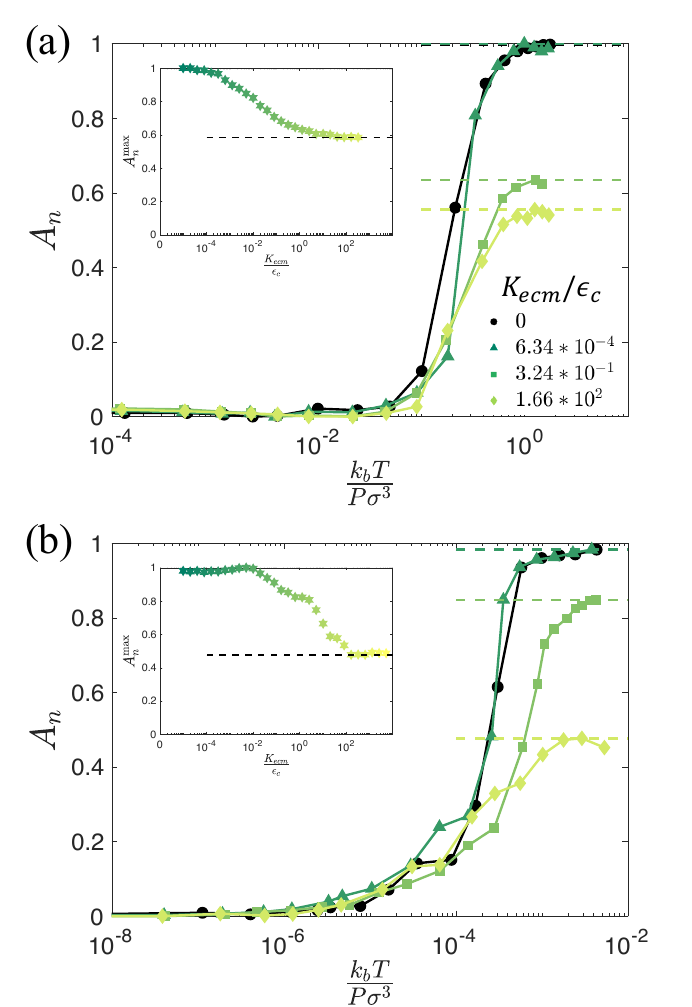}
\caption{$A_n$ is plotted versus $\frac{k_bT}{P\sigma^3}$ for cancer cells with (a) $\tau_p=25\sqrt{\frac{m\sigma^2}{\epsilon_c}}$ and $\beta=10^{-2}$ and (b) $\tau_p=2500\sqrt{\frac{m\sigma^2}{\epsilon_c}}$ and $\beta=10^{-3}$.  Stiffness of the ECM increases from dark to light green. Systems with $K_{ecm}=0$ are indicated by black filled circles. The dashed horizontal lines indicate $A_n^{\rm max}$ for different values of $K_{ecm}$. Insets: $A_n^{\rm max}$ is plotted as a function of $K_{ecm}/\epsilon_c$ for the data in the main panels. The black dashed horizontal lines indicate $A_n^{\rm max}$ in the $K_{ecm}/\epsilon_c \rightarrow \infty$ limit.}
  \label{fig:supression}
\end{figure}

In the previous section, we found that the degree of invasion $A_n$ is time-dependent, evolving toward the mixed (invaded) state for purely repulsive, passive cancer cells. Next, we examined how $A_n$ is affected by cancer cell cohesion and persistence. In Fig.~\ref{fig:transition} (b), we show that $A_n$ versus $k_b T/(P\sigma^3)$ for systems with attractive, active cancer cells does not possess strong time dependence. Moreover, the value of $A_n$ does not depend on the starting condition, reaching a steady-state value for times $t/\sqrt{m \sigma^2/\epsilon_c} < 5 \times 10^5$ as shown in the inset to Fig.~\ref{fig:transition} (b).  In Fig.~\ref{fig:effects} (a), we plot the steady-state $A_n$ versus $k_bT/(P\sigma^3)$ for three values of the attractive strength $\beta$ and five values of the persistence time $\tau_P$. $A_n$ is sigmoidal versus $k_b T/(P \sigma^3)$ for all $\beta$ and $\tau_P$ combinations studied. Thus, we can describe $A_n$ using 
\begin{equation}\label{fitA}
A_n = \frac{1}{2}\left(\textrm{tanh}\left[\textrm{log}_{10}\left(\frac{k_bT}{aP\sigma^3}\right)^b\right]+1\right),
\end{equation}
where $a(\beta,\tau_p)$ controls the value of $k_b T/(P\sigma^3)$ at which the rapid increase in $A_n$ occurs and $0.5 < b < 2.5$ controls the rate of this increase. We demonstrate high-quality fits of $A_n$ to Eq.~\ref{fitA} in Fig.~\ref{fig:effects} (a). Thus, when we plot $A_n$ as a function of $E_c^b$, where $E_c \equiv k_b T/(a P \sigma^3)$, the data collapses as shown in Fig.~\ref{fig:effects} (b).  We next determine the functional form for $a(\beta,\tau_p)$. First, $a$ for passive cancer cells ($\tau_p=0$) at a given $T$ and $P$ is similar to that for active cancer cells at small $\tau_p$. (See black/purple circles and black/purple diamonds in Fig.~\ref{fig:effects} (a).)  Next, $a$ decreases with increasing $\tau_p$ for large $\tau_p$, emphasizing that cancer cell persistence enhances invasion. Thus, a reasonable ansatz for $a$ at fixed $\beta$ is $a \sim 1/\tau_d$, where $\tau_d$ is the decorrelation time of the velocity autocorrelation function for the cancer cells,
\begin{equation}\label{Cvv}
    C_{vv}\left(t\right)=\left<\frac{\textbf{v}_q(t')\cdot\textbf{v}_q\left(t'+ t\right)}{\textbf{v}_q^2(t')}\right>_{q,t'}
\end{equation}
and 
\begin{equation}\label{td}
    C_{vv}\left(\tau_d\right)= e^{-1}.
\end{equation}
(See Fig.~\ref{fig:Cvv} (a).) To understand the dependence of $\tau_d$ on cancer cell packing fraction and $\tau_p$, we show $\tau_d$ for simplified systems composed of only repulsive, active cancer cells in periodic boundary conditions in Fig.~\ref{fig:Cvv} (b). $\tau_{d} \sim \tau_{\gamma}$ in the $\tau_p \rightarrow 0$ limit for all $\phi_c$, where $\tau_{\gamma}=\frac{\gamma}{m}$. For intermediate values of $\tau_p$, $\tau_d \sim \tau_p$, which agrees with the analytical result at $\phi_c=0$\cite{Caprini2021}:
\begin{equation}\label{Cvv0}
    C_{vv}\left(t\right)=\frac{\tau_{\gamma}e^{-\frac{t}{\tau_{\gamma}}}-\tau_p e^{-\frac{t}{\tau_p}}}{\tau_{\gamma}-\tau_p}.
\end{equation}
At large $\tau_p$, $\tau_d$ deviates from the analytical result as $\phi_c$ increases. In the inset to Fig.~\ref{fig:effects} (b), we verify that $a \sim \tau_{\gamma}/\tau_d$. 

At fixed $\tau_p$, $A_n$ shifts to larger $k_b T/(P\sigma^3)$ with increasing attraction strength $\beta \epsilon_c$ since cohesion makes it more difficult for cancer cells to invade the adipocytes. In the $\beta \rightarrow 0$ limit, $A_n$ converges to a time-dependent value for purely repulsive cancer cells. Thus, we expect $a \sim c_1(t) + c \beta \epsilon_c$, where $c_1(t)\sim k_bT_g/(P\sigma^3)$ in the long time limit and $a$ is not time-dependent when $c\beta \gg c_1(t)$. We show in Fig.~\ref{fig:effects} (b) that $a=c_1\left(1 +c_2 \beta \epsilon_c/(P\sigma^3) \right) \tau_{\gamma}/\tau_d$, where $c_1 \approx 0.06$ enforces $A_n=1/2$ at $E_c=1$ and $c_2 \approx 0.02$. 

\subsection*{Tethering of neighboring adipocytes}

The ECM surrounding the adipocytes provides structural support for adipose tissue. In this section, we investigate the effects of the ECM on breast cancer cell invasion. To model the structural support provided by the ECM, we add linear springs between the centers of mass, $\mathbf{R}_i=\frac{1}{N_v}\sum_{\mu=1}^{N_v}\mathbf{r}_{i\mu}$, of neighboring adipocytes $i$ and $j$. Adipocytes are considered neighbors when the distance between the centers of mass of adipocyte pairs is smaller than a threshold value $d_c$, such that the number of neighbors per adipocyte is $6$. The potential energy associated with the spring connecting adipocytes $i$ and $j$ is
\begin{equation}\label{Uecm}
    U_{ij}^{ECM} = \frac{K_{ecm}}{2}\left(\frac{R_{ij}-l_{ij0}}{\sigma}\right)^2,
\end{equation}
where $K_{ecm}/\sigma^2$ is the stiffness of the spring, $R_{ij}=|\mathbf{R}_i-\mathbf{R}_j|$, and $l_{ij0}$ is the equilibrium length, which is set equal to the adipocyte separations in the initial configuration. All vertices on adipocyte $i$ share the interaction with adipocyte $j$ equally, and thus the force on the vertex $\mu$ on adipocyte $i$ from adipocyte $j$ is given by 
\begin{equation}\label{Fecm}
    \mathbf{F}_{i\mu j}^{ECM} = -\frac{1}{N_v}\nabla_{\mathbf{R}_i} U_{ij}^{ECM}.
\end{equation}
In Fig.~\ref{fig:supression} (a) and (b), we calculate the interfacial area $A_n$ versus $k_b T/(P\sigma^3)$ for several values of $K_{ecm}$ and two values of $\tau_p$ and $\beta$. As discussed above, $A_n$ versus $k_bT/(P\sigma^3)$ is normalized between $0$ and $1$ using the interfacial area obtained for untethered ($K_{ecm}=0$) systems.  Tethering of adipocytes decreases the degree of invasion but not the onset value of $k_b T/(P\sigma^3)$ for invasion.  We find that the plateau values $A_n^{\rm max} < 1$ at large $k_b T/(P\sigma^3)$ for all $K_{ecm} >0$, but the values of $k_b T/(P\sigma^3)$ at which $A_n>0$ does not vary strongly with $K_{ecm}$.  In the insets to Fig.~\ref{fig:supression}, we show that $A_n^{\rm max}$ decreases monotonically from $1$ to lower values that depend on $\tau_p$ and $\beta$. Thus, the tethering of adipocytes by the ECM reduces the degree of cancer cell invasion into adipose tissue.

\subsection*{Packing fraction heterogeneity}

\begin{figure}[!htbp]
\centering
  \includegraphics[width=8cm]{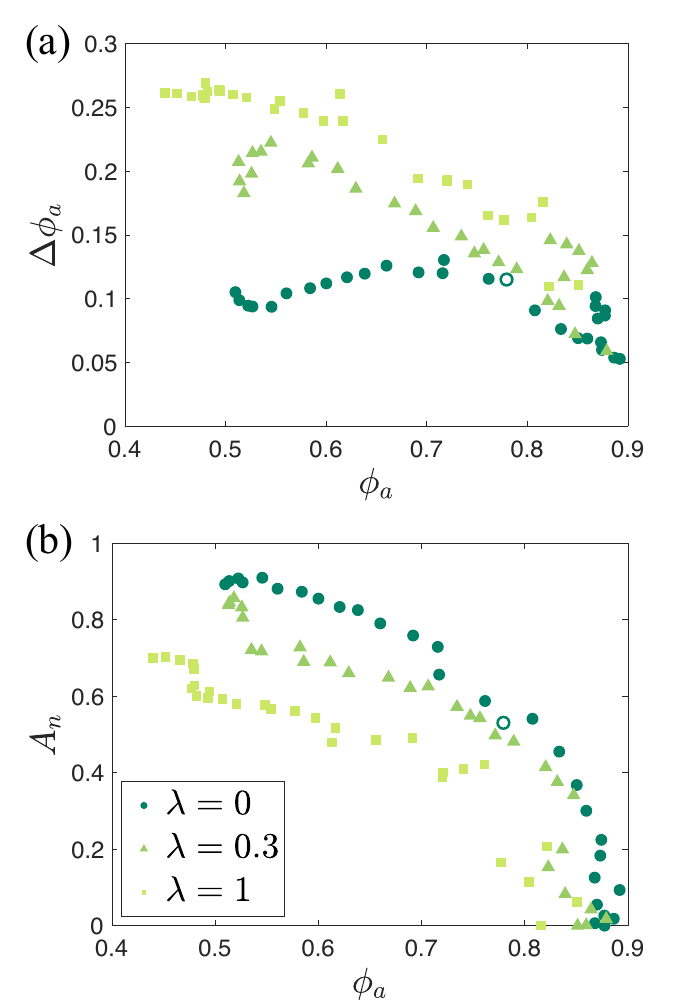}
  \caption{(a) The degree of invasion $A_n$ plotted versus the packing fraction of adipocytes $\phi_{a}$ at $\tau_P=25\sqrt{\frac{m\sigma^2}{\epsilon_c}}$, $\beta=10^{-2}$ and $k_bT/(P\sigma^3)=0.37$. $\lambda$ increases from dark to light green. (b) Standard deviation in the local packing fraction of adipocytes $\Delta \phi_{a}$ plotted versus $\phi_{a}$.  Open dark green circles indicate results for uniformly distributed adipocytes at a total packing fraction $\phi=0.72$.}
  \label{fig:variance}
\end{figure}

The mechanical properties of adipose tissue are affected by many factors including obesity, inflammation, and lipolysis.\cite{seo2015obesity,alsaggar2020silibinin,Zhu2022} Previous experimental studies of cancer cell migration through collagen networks demonstrated that geometrical confinement strongly affects cancer cell motion\cite{wolf2013physical,Ilina2020}. Thus, we expect the packing fraction of adipocytes to influence the degree of invasion. Adipocytes are initialized using athermal, quasistatic compression to reach a total packing fraction $\phi=0.72$ as described in the Methods section. We now vary the adipocyte packing fraction over the range $0.4 < \phi_a < 0.9$ by changing the length of the simulation box and shifting the positions of the cancer cells and adipocytes affinely, followed by energy minimization. We then add linear springs between neighboring adipocytes, setting $l_{ij0}=R_{ij}$ and $K_{ecm}=0.04\epsilon_c$, and calculate $A_n$ versus $\phi_a$ for $\tau_p/\sqrt{\frac{m\sigma^2}{\epsilon_c}}=25$ and $\beta=10^{-2}$. In Fig.~\ref{fig:variance} (a), we confirm that $A_n$ decreases with increasing adipocyte packing fraction and vanishes for $\phi_a \gtrsim 0.85$.   

In addition to packing fraction, the local structural and mechanical properties of adipose tissue are often heterogeneous.\cite{Alkhouli2013} We model structural and mechanical heterogeneity in adipose tissue by resetting $l_{ij0}$ of the tethering springs for half of the adipocytes in the $z$-direction to $l_{ij0}= (1-\lambda) R_{ij}$ and to $l_{ij0}=(1+\lambda)R_{ij}$ for the other half of the adipocytes.  In this idealized system, half of the adipocytes are tightly coupled with smaller equilibrium lengths, and half are loosely coupled with larger equilibrium lengths. This heterogeneity in equilibrium lengths gives rise to root-mean-square fluctuations in the local packing fraction that increase with $\lambda$. As shown in Fig.~\ref{fig:variance} (b), the maximum values of $\Delta \phi$ are $\approx 0.1$, $0.2$, and $0.25$ for $\lambda=0$, $0.3$, and $1$. In Fig.~\ref{fig:variance} (a), we show that invasion is less pronounced for heterogeneous systems compared to more uniform systems at the same adipocyte packing fraction. This result can be explained given that for sufficiently large $\lambda$, tightly coupled adipocytes exhibit smaller gaps between adipocytes that are less than the diameter of the cancer cells, reducing the degree of invasion in half of the system. Since the adipocytes are tethered, cancer cells invading the loosely coupled adipocytes do not affect invasion in the tightly coupled half of the system.

\section*{Conclusions and future directions}

In this work, we aimed to understand how the physical properties of cancer cells and adipose tissue influence breast cancer invasion. To this end, we developed discrete element method simulations of tumor invasion, modeling the cancer cells as cohesive, active, soft particles and the adipocytes as deformable, polyhedral particles. We initialized the system in a de-mixed state and quantified the degree of invasion as the interfacial area $A_t$ shared between the cancer cells and adipocytes. $A_t$ is bounded by the area of the interface in the initial de-mixed state $A_t^{\rm min}$ and the total surface area of the adipocytes $A_t^{\rm max}$, and we normalized the interfacial area by these two values so that $0 < A_n < 1$. We then calculated $A_n$ as a function of temperature, pressure, and the cohesive strength and persistence of cancer cell motion. We showed that $A_n$ can be collapsed when plotted as a function of a dimensionless energy scale $E_c$, where $E_c \ll 1$ indicates a de-mixed (non-invaded) state and $E_c \gg 1$ indicates a mixed (invaded) state. We found that $E_c$ increases with the mean-square fluctuations and persistence of cancer cell velocities and is inversely related to cancer cell cohesion and pressure in the system. We also investigated the physical effects of the extracellular matrix on cancer cell invasion. We demonstrated that tethering of neighboring adipocytes and denser packing of adipocytes inhibit invasion. In addition, we showed that heterogeneity in the local packing fraction for tethered adipocytes decreases $A_n$ relative to more uniform local packing.   

The current computational studies represent a first step toward modeling breast cancer invasion into adipose tissue as a physical process.  However, there are numerous important future research directions to pursue. First, we accounted for some of the effects of the ECM by tethering neighboring adipocytes to each other.  However, alignment of ECM fibers also strongly influences the speed and persistence of cancer cell velocities during invasion.\cite{wolf2013physical,Wang2018} In preliminary experimental studies, we have embedded plastic spherical beads within collagen matrices and quantified the local density and alignment of the collagen fibers, as well as the positions of the spherical beads. These studies have varied the bead packing fraction and collagen density and correlated enhanced cancer cell motion to collagen fiber alignment.\cite{sun2018novel} In future computational studies, we can incorporate separate fields into the simulations to account for local variations in collagen density and alignment, couple these fields to cancer cell velocities, and quantitatively compare simulated trajectories of the cancer cells with experimental results from \textit{in vitro}  systems where collagen fiber alignment can be controlled. Second, in the current simulations, we modeled cancer cells as active, soft spherical particles that do not change shape. However, we know that cancer cells can significantly alter their shapes to squeeze through narrow gaps and their motion has been correlated with shape elongation.\cite{Rahman2023, wolf2013physical,grosser2021cell} Thus, in future studies, we aim to model cancer cells using active, {\it deformable} particles. Studies have also suggested that breast cancer invasion into adipose tissue can induce lipolysis in adipocytes.\cite{Zhu2022} To model this effect, we can allow the equilibrium volume $v_{i0}$ of adipocyte $i$ to decrease over time when in contact with cancer cells. 

Finally, in our computational studies, we assumed that the cancer cell invasion time scale $\tau_{\rm inv}$ is shorter than the timescale of the cell cycle $\tau_{\rm cell}$, and thus we did not model cell growth, division, and apoptosis. As a result, de-mixed (non-invaded) systems in our current studies could become mixed (invaded) due to cancer cell proliferation on time scales longer than than $\tau_{\rm cell}$. Moreover, recent studies demonstrated that coupling cell growth rate and local pressure generates an interface instability with a fingering invasion pattern even when cancer cells do not migrate.\cite{Ye2024} This work emphasizes that systems with $\tau_{\rm cell} \ll \tau_{\rm inv}$ are also relevant to cancer invasion. In future computational studies, we will model cancer cell growth, division, and apoptosis and determine the mechanism of cancer cell invasion and shape of the invasion front as a function of $\tau_{\rm cell}/\tau_{\rm inv}$. 

\section*{Appendix}

\subsection*{Appendix A}
\begin{figure*}[!htbp]
\centering
  \includegraphics[width=17cm]{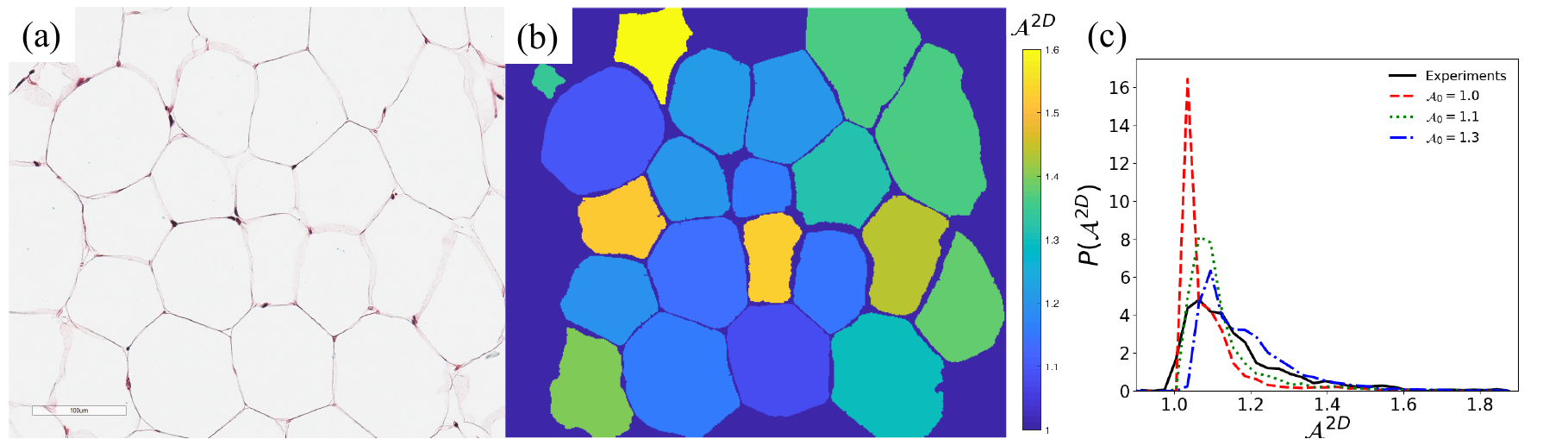}
\caption{(a) A representative histology slice of adipose tissue from a C57BL6 mouse. (b) Cell boundaries are identified through standard image processing techniques. Cells are colored by ${\cal A}_{2D} = p^2/(4\pi s)$, where $p$ is the perimeter and $s$ is the area. (c) Probability distribution of ${\cal A}_{2D}$ $P({\cal A}_{2D})$ for random 2D cross-sections of adipocytes in 3D from DEM invasion simulations with reference shape parameters ${\cal A}_0 = 1.0$ (dashed line), $1.1$ (dotted line), and $1.3$ (dot-dashed line) and from histological slices. The experimental data includes $\sim 1000$ adipocytes collected from histology samples from $5$ mice.}
  \label{fig:SP}
\end{figure*}

Due to the highly scattering nature of lipids, in situ 3D visualization of adipose tissue remains challenging. Instead, the extent of breast cancer invasion into adipose tissue is routinely assessed using 2D histology slices of tumors as shown in Fig.~\ref{fig:model} (a) and (b). We analyzed $\sim 1000$ adipocytes from five different mice with mammary tumors 11 days after implantation. For each slice, we identify the adipocytes and calculate the shape parameter of the cross-sections, $\mathcal{A}^{2D}=p^2/(4\pi s)$, where $p$ is the perimeter and $s$ is the area of the cross-section. In Fig.~\ref{fig:SP}, we compare the distribution $P(\mathcal{A}^{2D})$ from the 2D histology slices to the shape parameter distributions from random 2D slices through adipocytes from the DEM invasion simulations at several reference shape parameters ${\cal A}_0$ in 3D. We find that using ${\cal A}_0 \approx 1.1$ gives reasonable agreement for $P({\cal A}^{2D})$ between the experiments and simulations. 

\subsection*{Appendix B}
\begin{figure}[!htbp]
\centering
  \includegraphics[width=8cm]{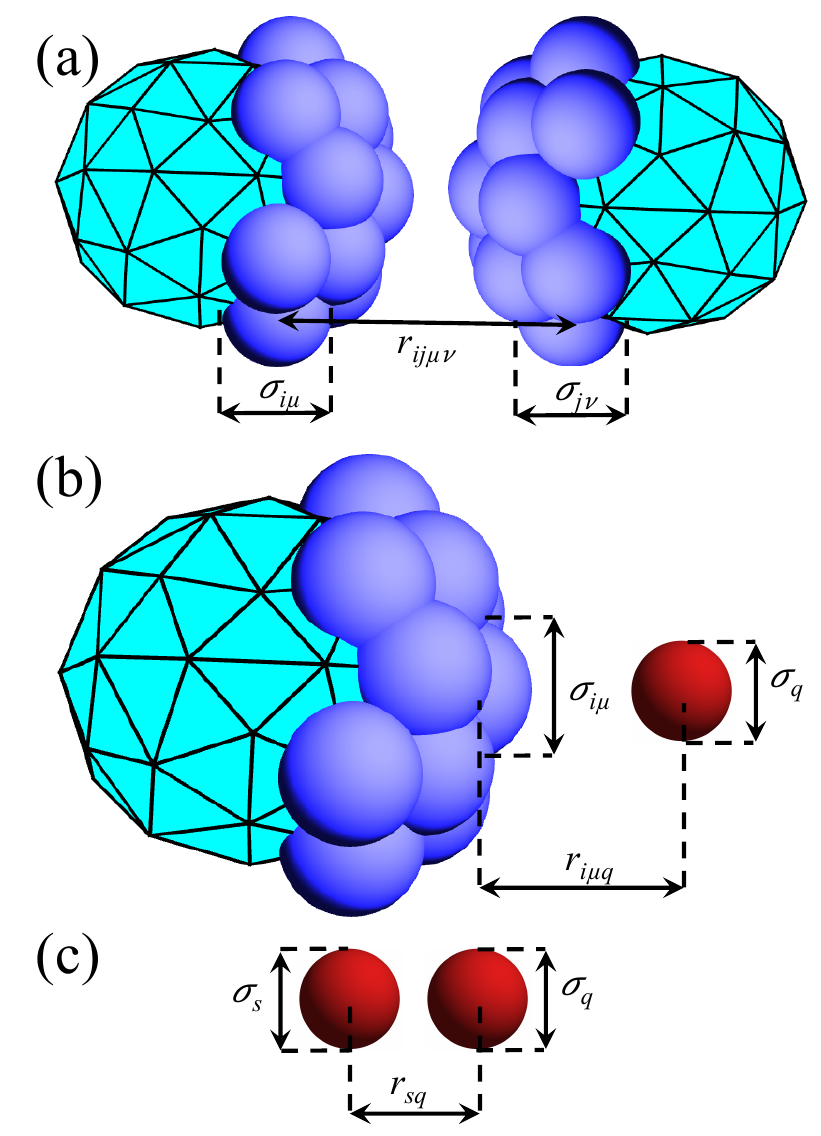}
\caption{(a) Distance $r_{ij\mu \nu}$ between vertex $\mu$ on adipocyte $i$ and vertex $\nu$ on adipocyte $j$. The average diameter of the vertices is $\sigma_{ij \mu \nu} = (\sigma_{i\mu}+\sigma_{j\nu})/2$. (b) Distance $r_{i\mu q}$ between vertex $\mu$ on adipocyte $i$ (with diameter $\sigma_{i\mu}$) and cancer cell $q$ (with diameter $\sigma_q$). (c) Distance $r_{qs}$ between cancer cells $q$ and $s$.}
  \label{fig:distance}
\end{figure}

In Eqs.~\ref{F1}-\ref{F2} in the Methods section, we define the pair forces between vertices on contacting adipocytes, between a vertex on an adipocyte and a contacting cancer cell, and between contacting cancer cells. Key parameters in these expressions include the separations $r_{ij\mu\nu}$, $r_{i\mu q}$, and $r_{qs}$ and the diameters $\sigma_{i\mu}$ and $\sigma_q$, which are pictured in Fig.~\ref{fig:distance}.    

\subsection*{Appendix C}\label{}

\begin{figure*}[ht]
\centering
  \includegraphics[width=17cm]{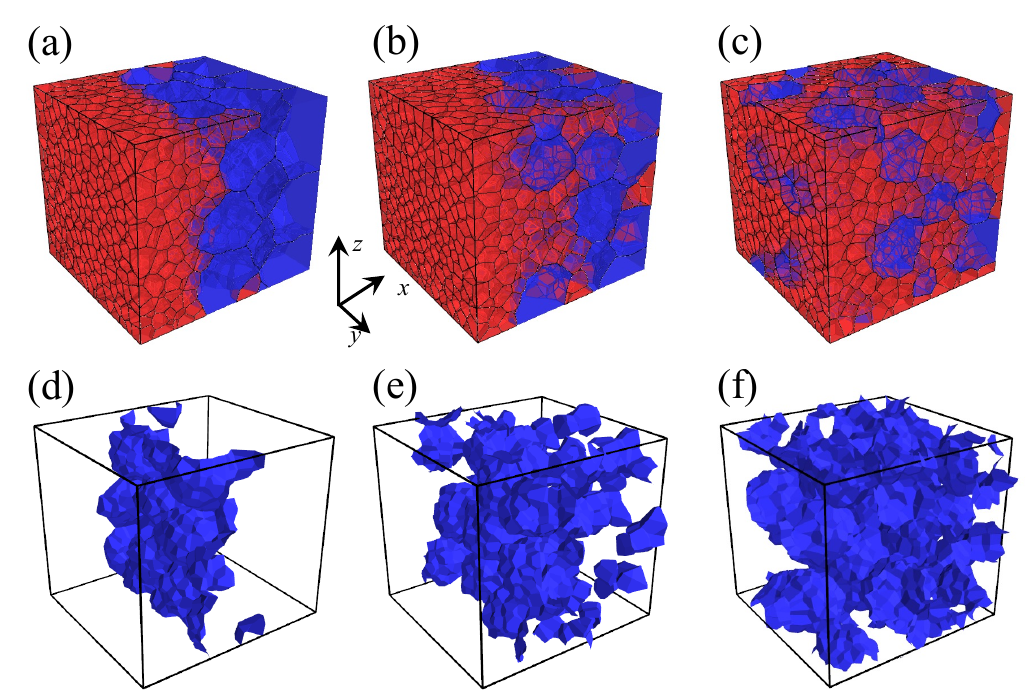}
\caption{(a)-(c) Voronoi tessellations of the centers of cancer cells (red shaded Voronoi polyhedra) and adipocytes (blue shaded Voronoi polyhedra) during cancer cell invasion into adipose tissue at several values of the shared interfacial area $A_t/\sigma^2$: (a) $610$, (b) $1280$, and (c) $1720$. (d)-(f) The Voronoi polyhedral faces that are shared between the cancer cells and adipocytes for panels (a)-(c). To improve visualization, we do not include faces of cancer cell Voronoi polyhedra that form the physical boundaries in the $x$-direction.}
  \label{fig:VT}
\end{figure*}

In this Appendix, we describe how we calculate the interfacial area $A_{t}$ shared by cancer cells and adipocytes and how we define the normalized interfacial area $0 < A_n < 1$. For a given configuration of cancer cells, adipocytes, and locations of the confining walls in the $x$-direction, we perform Voronoi tessellation of the positions of the cancer cells and vertices of the adipocytes, as shown in Fig.~\ref{fig:VT} (a)-(c). Next, we find all of the Voronoi polyhedral faces that are shared by a Voronoi polyhedron associated with a cancer cell and by a Voronoi polyhedron associated with a vertex of an adipocyte—as well as all of the faces of cancer cell Voronoi polyhedra that form the physical walls as shown in Fig.~\ref{fig:VT} (d)-(f). The interfacial area (between cancer cells and adipocytes and between cancer cells and the physical boundaries) $A_t$ is defined as the sum of the areas of these faces. The maximum value of $A_t$ is $A_t^{\rm max}=2L_f^2+\Sigma_{i=1}^{N_a} S_i$, where $S_i$ is the surface area of the Voronoi polyhedron for adipocyte $i$. The relative interfacial area $A_t/A_t^{\rm max}$ serves as an indicator of the degree of cancer cell invasion.

Note that the volume of the simulation box, as well as $A_t$ and $A_t^{\rm max}$, can increase with $k_bT/(P\sigma^3)$, especially at large values when $k_b T$ overcomes the cohesive energy $\beta \epsilon_c$. In this regime, the ratio $A_t/A_t^{\rm max}$ can vary with $k_bT/(P\sigma^3)$. Since we are not interested in changes in the relative interfacial area $A_t/A_t^{\rm max}$ caused by dilation of the simulation box, we define $rA_t=(A_t-\Delta A)/(A_t^{\rm max}-\Delta A)$, where $\Delta A=A_t^{\rm max}-\min(A_t^{\rm max})$ and $\min(A_t^{\rm max})$ is the value of $A_t^{\rm max}$ at which dilation occurs. We then define the normalized value $A_{n} = (rA_t-\min(rA_t))/(\max(rA_t)-\min(rA_t))$, where $\min(rA_t)$ and $\max(rA_t)$ are the minimum and maximum values of $rA_t$, such that $0 < A_n < 1$.

\section*{Acknowledgements}
We acknowledge support from National Institutes of Health R01 CA276392 (Y.Z., D.W., and C.S.O.), National Science Foundation DGE1650441 (G.B.), National Cancer Institute F31CA278410 (G.B.), National Cancer Institute R01CA259195 (C.F.) and through the Center on the Physics of Cancer Metabolism 1U54CA210184 (C.F.). This work was also supported by the High Performance Computing facilities operated by Yale’s Center for Research Computing.
\section*{Author Declarations}
\subsection*{Conflict of Interest}
The authors have no conflicts to disclose.
\subsection*{Ethics Approval}
Ethics approval is not required.
\subsection*{Author Contributions}
\textbf{Yitong Zheng}: Conceptualization (equal); Data curation (lead); Formal analysis (lead); Investigation (lead); Methodology (equal); Project Administration (equal); Software (lead); Validation (lead); Visualization (lead); Writing - original draft (equal); Writing - review \& editing (equal). \textbf{Dong Wang}: Conceptualization (supporting); Formal Analysis (supporting); Investigation (supporting); Methodology (supporting); Software (supporting); Validation (supporting); Visualization (supporting); Writing - original draft (supporting); Writing - review \& editing (supporting). \textbf{Garrett Beeghly}: Conceptualization (supporting); Data curation (supporting); Formal Analysis (supporting); Funding acquisition (supporing); Investigation (supporting); Writing - review \& editing (supporting). \textbf{Claudia Fischbach}: Conceptualization (equal); Funding acquisition (equal); Resources (supporting); Supervision (supporting); Writing - review \& editing (supporting). \textbf{Mark D. Shattuck}: Conceptualization (supporting); Formal Analysis (supporting); Methodology (supporting); Supervision (supporting). \textbf{Corey S. O'Hern}: Conceptualization (equal); Formal Analysis (supporting); Funding acquisition (equal); Methodology (equal); Project Administration (equal); Resources (lead); Supervision (lead); Writing - original draft (equal); Writing - review \& editing (equal).

\section*{Data Availability}
The data that support the findings of this study are available from the corresponding author upon reasonable request.
\section*{References}
\bibliography{mainFile}

\begin{thebibliography}{35}%
\makeatletter
\providecommand \@ifxundefined [1]{%
 \@ifx{#1\undefined}
}%
\providecommand \@ifnum [1]{%
 \ifnum #1\expandafter \@firstoftwo
 \else \expandafter \@secondoftwo
 \fi
}%
\providecommand \@ifx [1]{%
 \ifx #1\expandafter \@firstoftwo
 \else \expandafter \@secondoftwo
 \fi
}%
\providecommand \natexlab [1]{#1}%
\providecommand \enquote  [1]{``#1''}%
\providecommand \bibnamefont  [1]{#1}%
\providecommand \bibfnamefont [1]{#1}%
\providecommand \citenamefont [1]{#1}%
\providecommand \href@noop [0]{\@secondoftwo}%
\providecommand \href [0]{\begingroup \@sanitize@url \@href}%
\providecommand \@href[1]{\@@startlink{#1}\@@href}%
\providecommand \@@href[1]{\endgroup#1\@@endlink}%
\providecommand \@sanitize@url [0]{\catcode `\\12\catcode `\$12\catcode `\&12\catcode `\#12\catcode `\^12\catcode `\_12\catcode `\%12\relax}%
\providecommand \@@startlink[1]{}%
\providecommand \@@endlink[0]{}%
\providecommand \url  [0]{\begingroup\@sanitize@url \@url }%
\providecommand \@url [1]{\endgroup\@href {#1}{\urlprefix }}%
\providecommand \urlprefix  [0]{URL }%
\providecommand \Eprint [0]{\href }%
\providecommand \doibase [0]{http://dx.doi.org/}%
\providecommand \selectlanguage [0]{\@gobble}%
\providecommand \bibinfo  [0]{\@secondoftwo}%
\providecommand \bibfield  [0]{\@secondoftwo}%
\providecommand \translation [1]{[#1]}%
\providecommand \BibitemOpen [0]{}%
\providecommand \bibitemStop [0]{}%
\providecommand \bibitemNoStop [0]{.\EOS\space}%
\providecommand \EOS [0]{\spacefactor3000\relax}%
\providecommand \BibitemShut  [1]{\csname bibitem#1\endcsname}%
\let\auto@bib@innerbib\@empty
\bibitem [{\citenamefont {Majidpoor}\ and\ \citenamefont {Mortezaee}(2021)}]{majidpoor2021steps}%
  \BibitemOpen
  \bibfield  {author} {\bibinfo {author} {\bibfnamefont {J.}~\bibnamefont {Majidpoor}}\ and\ \bibinfo {author} {\bibfnamefont {K.}~\bibnamefont {Mortezaee}},\ }\bibfield  {title} {\enquote {\bibinfo {title} {Steps in metastasis: {A}n updated review},}\ }\href@noop {} {\bibfield  {journal} {\bibinfo  {journal} {Medical Oncology}\ }\textbf {\bibinfo {volume} {38}},\ \bibinfo {pages} {1--17} (\bibinfo {year} {2021})}\BibitemShut {NoStop}%
\bibitem [{\citenamefont {Novikov}\ \emph {et~al.}(2021)\citenamefont {Novikov}, \citenamefont {Zolotaryova}, \citenamefont {Gautreau},\ and\ \citenamefont {Denisov}}]{novikov2021mutational}%
  \BibitemOpen
  \bibfield  {author} {\bibinfo {author} {\bibfnamefont {N.~M.}\ \bibnamefont {Novikov}}, \bibinfo {author} {\bibfnamefont {S.~Y.}\ \bibnamefont {Zolotaryova}}, \bibinfo {author} {\bibfnamefont {A.~M.}\ \bibnamefont {Gautreau}}, \ and\ \bibinfo {author} {\bibfnamefont {E.~V.}\ \bibnamefont {Denisov}},\ }\bibfield  {title} {\enquote {\bibinfo {title} {Mutational drivers of cancer cell migration and invasion},}\ }\href@noop {} {\bibfield  {journal} {\bibinfo  {journal} {British Journal of Cancer}\ }\textbf {\bibinfo {volume} {124}},\ \bibinfo {pages} {102--114} (\bibinfo {year} {2021})}\BibitemShut {NoStop}%
\bibitem [{\citenamefont {Milotti}, \citenamefont {Stella},\ and\ \citenamefont {Chignola}(2017)}]{milotti2017pulsation}%
  \BibitemOpen
  \bibfield  {author} {\bibinfo {author} {\bibfnamefont {E.}~\bibnamefont {Milotti}}, \bibinfo {author} {\bibfnamefont {S.}~\bibnamefont {Stella}}, \ and\ \bibinfo {author} {\bibfnamefont {R.}~\bibnamefont {Chignola}},\ }\bibfield  {title} {\enquote {\bibinfo {title} {Pulsation-limited oxygen diffusion in the tumour microenvironment},}\ }\href@noop {} {\bibfield  {journal} {\bibinfo  {journal} {Scientific Reports}\ }\textbf {\bibinfo {volume} {7}},\ \bibinfo {pages} {39762} (\bibinfo {year} {2017})}\BibitemShut {NoStop}%
\bibitem [{\citenamefont {Marti}(2005)}]{marti2005angiogenesis}%
  \BibitemOpen
  \bibfield  {author} {\bibinfo {author} {\bibfnamefont {H.~H.}\ \bibnamefont {Marti}},\ }\bibfield  {title} {\enquote {\bibinfo {title} {Angiogenesis—{A} self-adapting principle in hypoxia},}\ \ }(\bibinfo  {publisher} {Springer},\ \bibinfo {address} {Basel},\ \bibinfo {year} {2005})\ pp.\ \bibinfo {pages} {163--180}\BibitemShut {NoStop}%
\bibitem [{\citenamefont {Sauer}\ \emph {et~al.}(2023)\citenamefont {Sauer}, \citenamefont {Grosser}, \citenamefont {Shahryari}, \citenamefont {Hayn}, \citenamefont {Guo}, \citenamefont {Braun}, \citenamefont {Briest}, \citenamefont {Wolf}, \citenamefont {Aktas}, \citenamefont {Horn}, \citenamefont {Sack},\ and\ \citenamefont {Käs}}]{Sauer2023}%
  \BibitemOpen
  \bibfield  {author} {\bibinfo {author} {\bibfnamefont {F.}~\bibnamefont {Sauer}}, \bibinfo {author} {\bibfnamefont {S.}~\bibnamefont {Grosser}}, \bibinfo {author} {\bibfnamefont {M.}~\bibnamefont {Shahryari}}, \bibinfo {author} {\bibfnamefont {A.}~\bibnamefont {Hayn}}, \bibinfo {author} {\bibfnamefont {J.}~\bibnamefont {Guo}}, \bibinfo {author} {\bibfnamefont {J.}~\bibnamefont {Braun}}, \bibinfo {author} {\bibfnamefont {S.}~\bibnamefont {Briest}}, \bibinfo {author} {\bibfnamefont {B.}~\bibnamefont {Wolf}}, \bibinfo {author} {\bibfnamefont {B.}~\bibnamefont {Aktas}}, \bibinfo {author} {\bibfnamefont {L.}~\bibnamefont {Horn}}, \bibinfo {author} {\bibfnamefont {I.}~\bibnamefont {Sack}}, \ and\ \bibinfo {author} {\bibfnamefont {J.~A.}\ \bibnamefont {Käs}},\ }\bibfield  {title} {\enquote {\bibinfo {title} {Changes in tissue fluidity predict tumor aggressiveness in vivo},}\ }\href@noop {} {\bibfield  {journal} {\bibinfo  {journal} {Advanced Science}\ }\textbf {\bibinfo {volume} {10}} (\bibinfo {year}
  {2023})}\BibitemShut {NoStop}%
\bibitem [{\citenamefont {Zhu}\ \emph {et~al.}(2022)\citenamefont {Zhu}, \citenamefont {Zhu}, \citenamefont {Hepler}, \citenamefont {Zhang}, \citenamefont {Park}, \citenamefont {Gliniak}, \citenamefont {Henry}, \citenamefont {Crewe}, \citenamefont {Bu}, \citenamefont {Zhang}, \citenamefont {Zhao}, \citenamefont {Morley}, \citenamefont {Li}, \citenamefont {Kim}, \citenamefont {Strand}, \citenamefont {Deng}, \citenamefont {Robino}, \citenamefont {Varlamov}, \citenamefont {Gordillo}, \citenamefont {Kolonin}, \citenamefont {Kusminski}, \citenamefont {Gupta},\ and\ \citenamefont {Scherer}}]{Zhu2022}%
  \BibitemOpen
  \bibfield  {author} {\bibinfo {author} {\bibfnamefont {Q.}~\bibnamefont {Zhu}}, \bibinfo {author} {\bibfnamefont {Y.}~\bibnamefont {Zhu}}, \bibinfo {author} {\bibfnamefont {C.}~\bibnamefont {Hepler}}, \bibinfo {author} {\bibfnamefont {Q.}~\bibnamefont {Zhang}}, \bibinfo {author} {\bibfnamefont {J.}~\bibnamefont {Park}}, \bibinfo {author} {\bibfnamefont {C.}~\bibnamefont {Gliniak}}, \bibinfo {author} {\bibfnamefont {G.~H.}\ \bibnamefont {Henry}}, \bibinfo {author} {\bibfnamefont {C.}~\bibnamefont {Crewe}}, \bibinfo {author} {\bibfnamefont {D.}~\bibnamefont {Bu}}, \bibinfo {author} {\bibfnamefont {Z.}~\bibnamefont {Zhang}}, \bibinfo {author} {\bibfnamefont {S.}~\bibnamefont {Zhao}}, \bibinfo {author} {\bibfnamefont {T.}~\bibnamefont {Morley}}, \bibinfo {author} {\bibfnamefont {N.}~\bibnamefont {Li}}, \bibinfo {author} {\bibfnamefont {D.~S.}\ \bibnamefont {Kim}}, \bibinfo {author} {\bibfnamefont {D.}~\bibnamefont {Strand}}, \bibinfo {author} {\bibfnamefont {Y.}~\bibnamefont {Deng}}, \bibinfo {author}
  {\bibfnamefont {J.~J.}\ \bibnamefont {Robino}}, \bibinfo {author} {\bibfnamefont {O.}~\bibnamefont {Varlamov}}, \bibinfo {author} {\bibfnamefont {R.}~\bibnamefont {Gordillo}}, \bibinfo {author} {\bibfnamefont {M.~G.}\ \bibnamefont {Kolonin}}, \bibinfo {author} {\bibfnamefont {C.~M.}\ \bibnamefont {Kusminski}}, \bibinfo {author} {\bibfnamefont {R.~K.}\ \bibnamefont {Gupta}}, \ and\ \bibinfo {author} {\bibfnamefont {P.~E.}\ \bibnamefont {Scherer}},\ }\bibfield  {title} {\enquote {\bibinfo {title} {Adipocyte mesenchymal transition contributes to mammary tumor progression},}\ }\href@noop {} {\bibfield  {journal} {\bibinfo  {journal} {Cell Reports}\ }\textbf {\bibinfo {volume} {40}},\ \bibinfo {pages} {111362} (\bibinfo {year} {2022})}\BibitemShut {NoStop}%
\bibitem [{\citenamefont {Zhang}\ \emph {et~al.}(2018)\citenamefont {Zhang}, \citenamefont {Di~Martino}, \citenamefont {Bowman}, \citenamefont {Campbell}, \citenamefont {Baksh}, \citenamefont {Simon-Vermot}, \citenamefont {Kim}, \citenamefont {Haldeman}, \citenamefont {Mondal}, \citenamefont {Yong-Gonzales}, \citenamefont {Abu-Akeel}, \citenamefont {Merghoub}, \citenamefont {Jones}, \citenamefont {Zhu}, \citenamefont {Arora}, \citenamefont {E.~Ariyan}, \citenamefont {Birsoy}, \citenamefont {D.~Wolchok}, \citenamefont {S.~Panageas}, \citenamefont {Hollmann}, \citenamefont {Bravo-Cordero}, \citenamefont {White},\ and\ \citenamefont {Richard}}]{zhang2018adipocyte}%
  \BibitemOpen
  \bibfield  {author} {\bibinfo {author} {\bibfnamefont {M.}~\bibnamefont {Zhang}}, \bibinfo {author} {\bibfnamefont {J.~S.}\ \bibnamefont {Di~Martino}}, \bibinfo {author} {\bibfnamefont {R.~L.}\ \bibnamefont {Bowman}}, \bibinfo {author} {\bibfnamefont {N.~R.}\ \bibnamefont {Campbell}}, \bibinfo {author} {\bibfnamefont {S.~C.}\ \bibnamefont {Baksh}}, \bibinfo {author} {\bibfnamefont {T.}~\bibnamefont {Simon-Vermot}}, \bibinfo {author} {\bibfnamefont {I.~S.}\ \bibnamefont {Kim}}, \bibinfo {author} {\bibfnamefont {P.}~\bibnamefont {Haldeman}}, \bibinfo {author} {\bibfnamefont {C.}~\bibnamefont {Mondal}}, \bibinfo {author} {\bibfnamefont {V.}~\bibnamefont {Yong-Gonzales}}, \bibinfo {author} {\bibfnamefont {M.}~\bibnamefont {Abu-Akeel}}, \bibinfo {author} {\bibfnamefont {R.~T.}\ \bibnamefont {Merghoub}}, \bibinfo {author} {\bibfnamefont {D.}~\bibnamefont {Jones}}, \bibinfo {author} {\bibfnamefont {X.~G.}\ \bibnamefont {Zhu}}, \bibinfo {author} {\bibfnamefont {A.}~\bibnamefont {Arora}}, \bibinfo {author}
  {\bibfnamefont {C.}~\bibnamefont {E.~Ariyan}}, \bibinfo {author} {\bibfnamefont {K.}~\bibnamefont {Birsoy}}, \bibinfo {author} {\bibfnamefont {J.}~\bibnamefont {D.~Wolchok}}, \bibinfo {author} {\bibfnamefont {K.}~\bibnamefont {S.~Panageas}}, \bibinfo {author} {\bibfnamefont {T.}~\bibnamefont {Hollmann}}, \bibinfo {author} {\bibfnamefont {J.~J.}\ \bibnamefont {Bravo-Cordero}}, \bibinfo {author} {\bibfnamefont {M.}~\bibnamefont {White}}, \ and\ \bibinfo {author} {\bibnamefont {Richard}},\ }\bibfield  {title} {\enquote {\bibinfo {title} {Adipocyte-derived lipids mediate melanoma progression via fatp proteins},}\ }\href@noop {} {\bibfield  {journal} {\bibinfo  {journal} {Cancer Discovery}\ }\textbf {\bibinfo {volume} {8}},\ \bibinfo {pages} {1006--1025} (\bibinfo {year} {2018})}\BibitemShut {NoStop}%
\bibitem [{\citenamefont {Friedl}\ \emph {et~al.}(2012)\citenamefont {Friedl}, \citenamefont {Locker}, \citenamefont {Sahai},\ and\ \citenamefont {Segall}}]{Friedl2012}%
  \BibitemOpen
  \bibfield  {author} {\bibinfo {author} {\bibfnamefont {P.}~\bibnamefont {Friedl}}, \bibinfo {author} {\bibfnamefont {J.}~\bibnamefont {Locker}}, \bibinfo {author} {\bibfnamefont {E.}~\bibnamefont {Sahai}}, \ and\ \bibinfo {author} {\bibfnamefont {J.~E.}\ \bibnamefont {Segall}},\ }\bibfield  {title} {\enquote {\bibinfo {title} {Classifying collective cancer cell invasion},}\ }\href {\doibase 10.1038/ncb2548} {\bibfield  {journal} {\bibinfo  {journal} {Nature Cell Biology}\ }\textbf {\bibinfo {volume} {14}},\ \bibinfo {pages} {777--783} (\bibinfo {year} {2012})}\BibitemShut {NoStop}%
\bibitem [{\citenamefont {Janiszewska}, \citenamefont {Primi},\ and\ \citenamefont {Izard}(2020)}]{Janiszewska2020}%
  \BibitemOpen
  \bibfield  {author} {\bibinfo {author} {\bibfnamefont {M.}~\bibnamefont {Janiszewska}}, \bibinfo {author} {\bibfnamefont {M.~C.}\ \bibnamefont {Primi}}, \ and\ \bibinfo {author} {\bibfnamefont {T.}~\bibnamefont {Izard}},\ }\bibfield  {title} {\enquote {\bibinfo {title} {Cell adhesion in cancer: Beyond the migration of single cells},}\ }\href {\doibase 10.1074/jbc.REV119.007759} {\bibfield  {journal} {\bibinfo  {journal} {Journal of Biological Chemistry}\ }\textbf {\bibinfo {volume} {295}},\ \bibinfo {pages} {2495--2505} (\bibinfo {year} {2020})}\BibitemShut {NoStop}%
\bibitem [{\citenamefont {Li}\ \emph {et~al.}(2017)\citenamefont {Li}, \citenamefont {Hebert}, \citenamefont {Lee}, \citenamefont {Xing}, \citenamefont {Boussommier-Calleja}, \citenamefont {Hynes}, \citenamefont {Lauffenburger},\ and\ \citenamefont {Kamm}}]{li2017macrophage}%
  \BibitemOpen
  \bibfield  {author} {\bibinfo {author} {\bibfnamefont {R.}~\bibnamefont {Li}}, \bibinfo {author} {\bibfnamefont {J.~D.}\ \bibnamefont {Hebert}}, \bibinfo {author} {\bibfnamefont {T.~A.}\ \bibnamefont {Lee}}, \bibinfo {author} {\bibfnamefont {H.}~\bibnamefont {Xing}}, \bibinfo {author} {\bibfnamefont {A.}~\bibnamefont {Boussommier-Calleja}}, \bibinfo {author} {\bibfnamefont {R.~O.}\ \bibnamefont {Hynes}}, \bibinfo {author} {\bibfnamefont {D.~A.}\ \bibnamefont {Lauffenburger}}, \ and\ \bibinfo {author} {\bibfnamefont {R.~D.}\ \bibnamefont {Kamm}},\ }\bibfield  {title} {\enquote {\bibinfo {title} {Macrophage-secreted tnf$\alpha$ and tgf$\beta$1 influence migration speed and persistence of cancer cells in 3d tissue culture via independent pathways},}\ }\href@noop {} {\bibfield  {journal} {\bibinfo  {journal} {Cancer Research}\ }\textbf {\bibinfo {volume} {77}},\ \bibinfo {pages} {279--290} (\bibinfo {year} {2017})}\BibitemShut {NoStop}%
\bibitem [{\citenamefont {Krakhmal}\ \emph {et~al.}(2015)\citenamefont {Krakhmal}, \citenamefont {Zavyalova}, \citenamefont {Denisov}, \citenamefont {Vtorushin},\ and\ \citenamefont {Perelmuter}}]{Krakhmal2015}%
  \BibitemOpen
  \bibfield  {author} {\bibinfo {author} {\bibfnamefont {N.~V.}\ \bibnamefont {Krakhmal}}, \bibinfo {author} {\bibfnamefont {M.~V.}\ \bibnamefont {Zavyalova}}, \bibinfo {author} {\bibfnamefont {E.~V.}\ \bibnamefont {Denisov}}, \bibinfo {author} {\bibfnamefont {S.~V.}\ \bibnamefont {Vtorushin}}, \ and\ \bibinfo {author} {\bibfnamefont {V.~M.}\ \bibnamefont {Perelmuter}},\ }\bibfield  {title} {\enquote {\bibinfo {title} {Cancer invasion: Patterns and mechanisms},}\ }\href@noop {} {\bibfield  {journal} {\bibinfo  {journal} {Acta Naturae}\ }\textbf {\bibinfo {volume} {7}},\ \bibinfo {pages} {17--28} (\bibinfo {year} {2015})}\BibitemShut {NoStop}%
\bibitem [{\citenamefont {Gonçalves}\ and\ \citenamefont {Garcia-Aznar}(2021)}]{Gon2021}%
  \BibitemOpen
  \bibfield  {author} {\bibinfo {author} {\bibfnamefont {I.~G.}\ \bibnamefont {Gonçalves}}\ and\ \bibinfo {author} {\bibfnamefont {J.~M.}\ \bibnamefont {Garcia-Aznar}},\ }\bibfield  {title} {\enquote {\bibinfo {title} {Extracellular matrix density regulates the formation of tumour spheroids through cell migration},}\ }\href {\doibase 10.1371/journal.pcbi.1008764} {\bibfield  {journal} {\bibinfo  {journal} {PLOS Computational Biology}\ }\textbf {\bibinfo {volume} {17}},\ \bibinfo {pages} {e1008764} (\bibinfo {year} {2021})}\BibitemShut {NoStop}%
\bibitem [{\citenamefont {Seo}\ \emph {et~al.}(2015)\citenamefont {Seo}, \citenamefont {Bhardwaj}, \citenamefont {Choi}, \citenamefont {Gonzalez}, \citenamefont {Eguiluz}, \citenamefont {Wang}, \citenamefont {Mohanan}, \citenamefont {Morris}, \citenamefont {Du}, \citenamefont {Zhou}, \citenamefont {Vahdat}, \citenamefont {Verma}, \citenamefont {Elemento}, \citenamefont {Hudis}, \citenamefont {Williams}, \citenamefont {Gourdon}, \citenamefont {Dannenberg},\ and\ \citenamefont {Fischbach}}]{seo2015obesity}%
  \BibitemOpen
  \bibfield  {author} {\bibinfo {author} {\bibfnamefont {B.~R.}\ \bibnamefont {Seo}}, \bibinfo {author} {\bibfnamefont {P.}~\bibnamefont {Bhardwaj}}, \bibinfo {author} {\bibfnamefont {S.}~\bibnamefont {Choi}}, \bibinfo {author} {\bibfnamefont {J.}~\bibnamefont {Gonzalez}}, \bibinfo {author} {\bibfnamefont {R.~C.~A.}\ \bibnamefont {Eguiluz}}, \bibinfo {author} {\bibfnamefont {K.}~\bibnamefont {Wang}}, \bibinfo {author} {\bibfnamefont {S.}~\bibnamefont {Mohanan}}, \bibinfo {author} {\bibfnamefont {P.~G.}\ \bibnamefont {Morris}}, \bibinfo {author} {\bibfnamefont {B.}~\bibnamefont {Du}}, \bibinfo {author} {\bibfnamefont {X.~K.}\ \bibnamefont {Zhou}}, \bibinfo {author} {\bibfnamefont {L.~T.}\ \bibnamefont {Vahdat}}, \bibinfo {author} {\bibfnamefont {A.}~\bibnamefont {Verma}}, \bibinfo {author} {\bibfnamefont {O.}~\bibnamefont {Elemento}}, \bibinfo {author} {\bibfnamefont {C.~A.}\ \bibnamefont {Hudis}}, \bibinfo {author} {\bibfnamefont {R.~M.}\ \bibnamefont {Williams}}, \bibinfo {author} {\bibfnamefont
  {D.}~\bibnamefont {Gourdon}}, \bibinfo {author} {\bibfnamefont {A.~J.}\ \bibnamefont {Dannenberg}}, \ and\ \bibinfo {author} {\bibfnamefont {C.}~\bibnamefont {Fischbach}},\ }\bibfield  {title} {\enquote {\bibinfo {title} {Obesity-dependent changes in interstitial ecm mechanics promote breast tumorigenesis},}\ }\href@noop {} {\bibfield  {journal} {\bibinfo  {journal} {Science Translational Medicine}\ }\textbf {\bibinfo {volume} {7}},\ \bibinfo {pages} {301ra130} (\bibinfo {year} {2015})}\BibitemShut {NoStop}%
\bibitem [{\citenamefont {Shimpi}\ \emph {et~al.}(2023)\citenamefont {Shimpi}, \citenamefont {Williams}, \citenamefont {Ling}, \citenamefont {Tamir}, \citenamefont {White},\ and\ \citenamefont {Fischbach}}]{shimpi2023phosphoproteomic}%
  \BibitemOpen
  \bibfield  {author} {\bibinfo {author} {\bibfnamefont {A.~A.}\ \bibnamefont {Shimpi}}, \bibinfo {author} {\bibfnamefont {E.~D.}\ \bibnamefont {Williams}}, \bibinfo {author} {\bibfnamefont {L.}~\bibnamefont {Ling}}, \bibinfo {author} {\bibfnamefont {T.}~\bibnamefont {Tamir}}, \bibinfo {author} {\bibfnamefont {F.~M.}\ \bibnamefont {White}}, \ and\ \bibinfo {author} {\bibfnamefont {C.}~\bibnamefont {Fischbach}},\ }\bibfield  {title} {\enquote {\bibinfo {title} {Phosphoproteomic changes induced by cell-derived matrix and their effect on tumor cell migration and cytoskeleton remodeling},}\ }\href@noop {} {\bibfield  {journal} {\bibinfo  {journal} {ACS Biomaterials Science \& Engineering}\ }\textbf {\bibinfo {volume} {9}},\ \bibinfo {pages} {6835--6848} (\bibinfo {year} {2023})}\BibitemShut {NoStop}%
\bibitem [{\citenamefont {Wolf}\ \emph {et~al.}(2013)\citenamefont {Wolf}, \citenamefont {Te~Lindert}, \citenamefont {Krause}, \citenamefont {Alexander}, \citenamefont {Te~Riet}, \citenamefont {Willis}, \citenamefont {Hoffman}, \citenamefont {Figdor}, \citenamefont {Weiss},\ and\ \citenamefont {Friedl}}]{wolf2013physical}%
  \BibitemOpen
  \bibfield  {author} {\bibinfo {author} {\bibfnamefont {K.}~\bibnamefont {Wolf}}, \bibinfo {author} {\bibfnamefont {M.}~\bibnamefont {Te~Lindert}}, \bibinfo {author} {\bibfnamefont {M.}~\bibnamefont {Krause}}, \bibinfo {author} {\bibfnamefont {S.}~\bibnamefont {Alexander}}, \bibinfo {author} {\bibfnamefont {J.}~\bibnamefont {Te~Riet}}, \bibinfo {author} {\bibfnamefont {A.~L.}\ \bibnamefont {Willis}}, \bibinfo {author} {\bibfnamefont {R.~M.}\ \bibnamefont {Hoffman}}, \bibinfo {author} {\bibfnamefont {C.~G.}\ \bibnamefont {Figdor}}, \bibinfo {author} {\bibfnamefont {S.~J.}\ \bibnamefont {Weiss}}, \ and\ \bibinfo {author} {\bibfnamefont {P.}~\bibnamefont {Friedl}},\ }\bibfield  {title} {\enquote {\bibinfo {title} {Physical limits of cell migration: control by ecm space and nuclear deformation and tuning by proteolysis and traction force},}\ }\href@noop {} {\bibfield  {journal} {\bibinfo  {journal} {Journal of Cell Biology}\ }\textbf {\bibinfo {volume} {201}},\ \bibinfo {pages} {1069--1084} (\bibinfo {year}
  {2013})}\BibitemShut {NoStop}%
\bibitem [{\citenamefont {Ton}\ \emph {et~al.}(2024)\citenamefont {Ton}, \citenamefont {MacKeith}, \citenamefont {Shattuck},\ and\ \citenamefont {O'Hern}}]{ton2024mechanical}%
  \BibitemOpen
  \bibfield  {author} {\bibinfo {author} {\bibfnamefont {A.~T.}\ \bibnamefont {Ton}}, \bibinfo {author} {\bibfnamefont {A.~K.}\ \bibnamefont {MacKeith}}, \bibinfo {author} {\bibfnamefont {M.~D.}\ \bibnamefont {Shattuck}}, \ and\ \bibinfo {author} {\bibfnamefont {C.~S.}\ \bibnamefont {O'Hern}},\ }\bibfield  {title} {\enquote {\bibinfo {title} {Mechanical plasticity of cell membranes enhances epithelial wound closure},}\ }\href@noop {} {\bibfield  {journal} {\bibinfo  {journal} {Physical Review Research}\ }\textbf {\bibinfo {volume} {6}},\ \bibinfo {pages} {L012036} (\bibinfo {year} {2024})}\BibitemShut {NoStop}%
\bibitem [{\citenamefont {Wang}\ \emph {et~al.}(2021)\citenamefont {Wang}, \citenamefont {Treado}, \citenamefont {Boromand}, \citenamefont {Norwick}, \citenamefont {Murrell}, \citenamefont {Shattuck},\ and\ \citenamefont {O'Hern}}]{Wang2021}%
  \BibitemOpen
  \bibfield  {author} {\bibinfo {author} {\bibfnamefont {D.}~\bibnamefont {Wang}}, \bibinfo {author} {\bibfnamefont {J.~D.}\ \bibnamefont {Treado}}, \bibinfo {author} {\bibfnamefont {A.}~\bibnamefont {Boromand}}, \bibinfo {author} {\bibfnamefont {B.}~\bibnamefont {Norwick}}, \bibinfo {author} {\bibfnamefont {M.~P.}\ \bibnamefont {Murrell}}, \bibinfo {author} {\bibfnamefont {M.~D.}\ \bibnamefont {Shattuck}}, \ and\ \bibinfo {author} {\bibfnamefont {C.~S.}\ \bibnamefont {O'Hern}},\ }\bibfield  {title} {\enquote {\bibinfo {title} {The structural, vibrational, and mechanical properties of jammed packings of deformable particles in three dimensions},}\ }\href@noop {} {\bibfield  {journal} {\bibinfo  {journal} {Soft Matter}\ }\textbf {\bibinfo {volume} {17}},\ \bibinfo {pages} {9901--9915} (\bibinfo {year} {2021})}\BibitemShut {NoStop}%
\bibitem [{\citenamefont {Cheng}\ \emph {et~al.}(2022)\citenamefont {Cheng}, \citenamefont {Treado}, \citenamefont {Lonial}, \citenamefont {Habdas}, \citenamefont {Weeks}, \citenamefont {Shattuck},\ and\ \citenamefont {O'Hern}}]{Cheng2022}%
  \BibitemOpen
  \bibfield  {author} {\bibinfo {author} {\bibfnamefont {Y.}~\bibnamefont {Cheng}}, \bibinfo {author} {\bibfnamefont {J.~D.}\ \bibnamefont {Treado}}, \bibinfo {author} {\bibfnamefont {B.~F.}\ \bibnamefont {Lonial}}, \bibinfo {author} {\bibfnamefont {P.}~\bibnamefont {Habdas}}, \bibinfo {author} {\bibfnamefont {E.~R.}\ \bibnamefont {Weeks}}, \bibinfo {author} {\bibfnamefont {M.~D.}\ \bibnamefont {Shattuck}}, \ and\ \bibinfo {author} {\bibfnamefont {C.~S.}\ \bibnamefont {O'Hern}},\ }\bibfield  {title} {\enquote {\bibinfo {title} {Hopper flows of deformable particles},}\ }\href {\doibase 10.1039/d2sm01079h} {\bibfield  {journal} {\bibinfo  {journal} {Soft Matter}\ }\textbf {\bibinfo {volume} {18}},\ \bibinfo {pages} {8071--8086} (\bibinfo {year} {2022})}\BibitemShut {NoStop}%
\bibitem [{\citenamefont {Boromand}\ \emph {et~al.}(2018)\citenamefont {Boromand}, \citenamefont {Signoriello}, \citenamefont {Ye}, \citenamefont {O'Hern},\ and\ \citenamefont {Shattuck}}]{Boromand2018}%
  \BibitemOpen
  \bibfield  {author} {\bibinfo {author} {\bibfnamefont {A.}~\bibnamefont {Boromand}}, \bibinfo {author} {\bibfnamefont {A.}~\bibnamefont {Signoriello}}, \bibinfo {author} {\bibfnamefont {F.}~\bibnamefont {Ye}}, \bibinfo {author} {\bibfnamefont {C.~S.}\ \bibnamefont {O'Hern}}, \ and\ \bibinfo {author} {\bibfnamefont {M.~D.}\ \bibnamefont {Shattuck}},\ }\bibfield  {title} {\enquote {\bibinfo {title} {Jamming of deformable polygons},}\ }\href@noop {} {\bibfield  {journal} {\bibinfo  {journal} {Physical Review Letters}\ }\textbf {\bibinfo {volume} {121}},\ \bibinfo {pages} {248003} (\bibinfo {year} {2018})}\BibitemShut {NoStop}%
\bibitem [{\citenamefont {Treado}\ \emph {et~al.}(2022)\citenamefont {Treado}, \citenamefont {Roddy}, \citenamefont {Théroux-Rancourt}, \citenamefont {Zhang}, \citenamefont {Ambrose}, \citenamefont {Brodersen}, \citenamefont {Shattuck},\ and\ \citenamefont {O'Hern}}]{Treado2022}%
  \BibitemOpen
  \bibfield  {author} {\bibinfo {author} {\bibfnamefont {J.~D.}\ \bibnamefont {Treado}}, \bibinfo {author} {\bibfnamefont {A.~B.}\ \bibnamefont {Roddy}}, \bibinfo {author} {\bibfnamefont {G.}~\bibnamefont {Théroux-Rancourt}}, \bibinfo {author} {\bibfnamefont {L.}~\bibnamefont {Zhang}}, \bibinfo {author} {\bibfnamefont {C.}~\bibnamefont {Ambrose}}, \bibinfo {author} {\bibfnamefont {C.~R.}\ \bibnamefont {Brodersen}}, \bibinfo {author} {\bibfnamefont {M.~D.}\ \bibnamefont {Shattuck}}, \ and\ \bibinfo {author} {\bibfnamefont {C.~S.}\ \bibnamefont {O'Hern}},\ }\bibfield  {title} {\enquote {\bibinfo {title} {Localized growth and remodelling drives spongy mesophyll morphogenesis},}\ }\href@noop {} {\bibfield  {journal} {\bibinfo  {journal} {Journal of the Royal Society Interface}\ }\textbf {\bibinfo {volume} {19}},\ \bibinfo {pages} {20220602} (\bibinfo {year} {2022})}\BibitemShut {NoStop}%
\bibitem [{\citenamefont {Treado}\ \emph {et~al.}(2021)\citenamefont {Treado}, \citenamefont {Wang}, \citenamefont {Boromand}, \citenamefont {Murrell}, \citenamefont {Shattuck},\ and\ \citenamefont {O'Hern}}]{Treado2021}%
  \BibitemOpen
  \bibfield  {author} {\bibinfo {author} {\bibfnamefont {J.~D.}\ \bibnamefont {Treado}}, \bibinfo {author} {\bibfnamefont {D.}~\bibnamefont {Wang}}, \bibinfo {author} {\bibfnamefont {A.}~\bibnamefont {Boromand}}, \bibinfo {author} {\bibfnamefont {M.~P.}\ \bibnamefont {Murrell}}, \bibinfo {author} {\bibfnamefont {M.~D.}\ \bibnamefont {Shattuck}}, \ and\ \bibinfo {author} {\bibfnamefont {C.~S.}\ \bibnamefont {O'Hern}},\ }\bibfield  {title} {\enquote {\bibinfo {title} {Bridging particle deformability and collective response in soft solids},}\ }\href@noop {} {\bibfield  {journal} {\bibinfo  {journal} {Physical Review Materials}\ }\textbf {\bibinfo {volume} {5}},\ \bibinfo {pages} {055605} (\bibinfo {year} {2021})}\BibitemShut {NoStop}%
\bibitem [{\citenamefont {Manning}(2023)}]{manning2023essay}%
  \BibitemOpen
  \bibfield  {author} {\bibinfo {author} {\bibfnamefont {M.~L.}\ \bibnamefont {Manning}},\ }\bibfield  {title} {\enquote {\bibinfo {title} {Essay: Collections of deformable particles present exciting challenges for soft matter and biological physics},}\ }\href@noop {} {\bibfield  {journal} {\bibinfo  {journal} {Physical Review Letters}\ }\textbf {\bibinfo {volume} {130}},\ \bibinfo {pages} {130002} (\bibinfo {year} {2023})}\BibitemShut {NoStop}%
\bibitem [{\citenamefont {Parlee}\ \emph {et~al.}(2014)\citenamefont {Parlee}, \citenamefont {Lentz}, \citenamefont {Mori},\ and\ \citenamefont {MacDougald}}]{parlee2014quantifying}%
  \BibitemOpen
  \bibfield  {author} {\bibinfo {author} {\bibfnamefont {S.~D.}\ \bibnamefont {Parlee}}, \bibinfo {author} {\bibfnamefont {S.~I.}\ \bibnamefont {Lentz}}, \bibinfo {author} {\bibfnamefont {H.}~\bibnamefont {Mori}}, \ and\ \bibinfo {author} {\bibfnamefont {O.~A.}\ \bibnamefont {MacDougald}},\ }\bibfield  {title} {\enquote {\bibinfo {title} {Quantifying size and number of adipocytes in adipose tissue},}\ }in\ \href@noop {} {\emph {\bibinfo {booktitle} {Methods in Enzymology}}},\ Vol.\ \bibinfo {volume} {537}\ (\bibinfo  {publisher} {Elsevier},\ \bibinfo {address} {Cambridge},\ \bibinfo {year} {2014})\ pp.\ \bibinfo {pages} {93--122}\BibitemShut {NoStop}%
\bibitem [{\citenamefont {Truongvo}\ \emph {et~al.}(2017)\citenamefont {Truongvo}, \citenamefont {Kennedy}, \citenamefont {Chen}, \citenamefont {Chen}, \citenamefont {Berndt}, \citenamefont {Agarwal}, \citenamefont {Zhu}, \citenamefont {Nakshatri}, \citenamefont {Wallace}, \citenamefont {Na}, \citenamefont {Yokota},\ and\ \citenamefont {Ryu}}]{Truongvo2017}%
  \BibitemOpen
  \bibfield  {author} {\bibinfo {author} {\bibfnamefont {T.~N.}\ \bibnamefont {Truongvo}}, \bibinfo {author} {\bibfnamefont {R.~M.}\ \bibnamefont {Kennedy}}, \bibinfo {author} {\bibfnamefont {H.}~\bibnamefont {Chen}}, \bibinfo {author} {\bibfnamefont {A.}~\bibnamefont {Chen}}, \bibinfo {author} {\bibfnamefont {A.}~\bibnamefont {Berndt}}, \bibinfo {author} {\bibfnamefont {M.}~\bibnamefont {Agarwal}}, \bibinfo {author} {\bibfnamefont {L.}~\bibnamefont {Zhu}}, \bibinfo {author} {\bibfnamefont {H.}~\bibnamefont {Nakshatri}}, \bibinfo {author} {\bibfnamefont {J.}~\bibnamefont {Wallace}}, \bibinfo {author} {\bibfnamefont {S.}~\bibnamefont {Na}}, \bibinfo {author} {\bibfnamefont {H.}~\bibnamefont {Yokota}}, \ and\ \bibinfo {author} {\bibfnamefont {J.~E.}\ \bibnamefont {Ryu}},\ }\bibfield  {title} {\enquote {\bibinfo {title} {Microfluidic channel for characterizing normal and breast cancer cells},}\ }\href@noop {} {\bibfield  {journal} {\bibinfo  {journal} {Journal of Micromechanics and Microengineering}\ }\textbf
  {\bibinfo {volume} {27}},\ \bibinfo {pages} {035017} (\bibinfo {year} {2017})}\BibitemShut {NoStop}%
\bibitem [{\citenamefont {Debets}, \citenamefont {Wit},\ and\ \citenamefont {Janssen}(2021)}]{Debets2021}%
  \BibitemOpen
  \bibfield  {author} {\bibinfo {author} {\bibfnamefont {V.~E.}\ \bibnamefont {Debets}}, \bibinfo {author} {\bibfnamefont {X.~M.~D.}\ \bibnamefont {Wit}}, \ and\ \bibinfo {author} {\bibfnamefont {L.~M.}\ \bibnamefont {Janssen}},\ }\bibfield  {title} {\enquote {\bibinfo {title} {Cage length controls the nonmonotonic dynamics of active glassy matter},}\ }\href@noop {} {\bibfield  {journal} {\bibinfo  {journal} {Physical Review Letters}\ }\textbf {\bibinfo {volume} {127}} (\bibinfo {year} {2021})}\BibitemShut {NoStop}%
\bibitem [{\citenamefont {Shaebani}\ \emph {et~al.}(2020)\citenamefont {Shaebani}, \citenamefont {Wysocki}, \citenamefont {Winkler}, \citenamefont {Gompper},\ and\ \citenamefont {Rieger}}]{shaebani2020computational}%
  \BibitemOpen
  \bibfield  {author} {\bibinfo {author} {\bibfnamefont {M.~R.}\ \bibnamefont {Shaebani}}, \bibinfo {author} {\bibfnamefont {A.}~\bibnamefont {Wysocki}}, \bibinfo {author} {\bibfnamefont {R.~G.}\ \bibnamefont {Winkler}}, \bibinfo {author} {\bibfnamefont {G.}~\bibnamefont {Gompper}}, \ and\ \bibinfo {author} {\bibfnamefont {H.}~\bibnamefont {Rieger}},\ }\bibfield  {title} {\enquote {\bibinfo {title} {Computational models for active matter},}\ }\href@noop {} {\bibfield  {journal} {\bibinfo  {journal} {Nature Reviews Physics}\ }\textbf {\bibinfo {volume} {2}},\ \bibinfo {pages} {181--199} (\bibinfo {year} {2020})}\BibitemShut {NoStop}%
\bibitem [{\citenamefont {Wang}\ \emph {et~al.}(2018)\citenamefont {Wang}, \citenamefont {Pearson}, \citenamefont {Kutys}, \citenamefont {Choi}, \citenamefont {Wozniak}, \citenamefont {Baker},\ and\ \citenamefont {Chen}}]{Wang2018}%
  \BibitemOpen
  \bibfield  {author} {\bibinfo {author} {\bibfnamefont {W.~Y.}\ \bibnamefont {Wang}}, \bibinfo {author} {\bibfnamefont {A.~T.}\ \bibnamefont {Pearson}}, \bibinfo {author} {\bibfnamefont {M.~L.}\ \bibnamefont {Kutys}}, \bibinfo {author} {\bibfnamefont {C.~K.}\ \bibnamefont {Choi}}, \bibinfo {author} {\bibfnamefont {M.~A.}\ \bibnamefont {Wozniak}}, \bibinfo {author} {\bibfnamefont {B.~M.}\ \bibnamefont {Baker}}, \ and\ \bibinfo {author} {\bibfnamefont {C.~S.}\ \bibnamefont {Chen}},\ }\bibfield  {title} {\enquote {\bibinfo {title} {Extracellular matrix alignment dictates the organization of focal adhesions and directs uniaxial cell migration},}\ }\href@noop {} {\bibfield  {journal} {\bibinfo  {journal} {APL Bioengineering}\ }\textbf {\bibinfo {volume} {2}},\ \bibinfo {pages} {046107} (\bibinfo {year} {2018})}\BibitemShut {NoStop}%
\bibitem [{\citenamefont {Caprini}\ and\ \citenamefont {Marconi}(2021)}]{Caprini2021}%
  \BibitemOpen
  \bibfield  {author} {\bibinfo {author} {\bibfnamefont {L.}~\bibnamefont {Caprini}}\ and\ \bibinfo {author} {\bibfnamefont {U.~M.~B.}\ \bibnamefont {Marconi}},\ }\bibfield  {title} {\enquote {\bibinfo {title} {Inertial self-propelled particles},}\ }\href@noop {} {\bibfield  {journal} {\bibinfo  {journal} {Journal of Chemical Physics}\ }\textbf {\bibinfo {volume} {154}},\ \bibinfo {pages} {024902} (\bibinfo {year} {2021})}\BibitemShut {NoStop}%
\bibitem [{\citenamefont {Alsaggar}\ \emph {et~al.}(2020)\citenamefont {Alsaggar}, \citenamefont {Bdour}, \citenamefont {Ababneh}, \citenamefont {El-Elimat}, \citenamefont {Qinna},\ and\ \citenamefont {Alzoubi}}]{alsaggar2020silibinin}%
  \BibitemOpen
  \bibfield  {author} {\bibinfo {author} {\bibfnamefont {M.}~\bibnamefont {Alsaggar}}, \bibinfo {author} {\bibfnamefont {S.}~\bibnamefont {Bdour}}, \bibinfo {author} {\bibfnamefont {Q.}~\bibnamefont {Ababneh}}, \bibinfo {author} {\bibfnamefont {T.}~\bibnamefont {El-Elimat}}, \bibinfo {author} {\bibfnamefont {N.}~\bibnamefont {Qinna}}, \ and\ \bibinfo {author} {\bibfnamefont {K.~H.}\ \bibnamefont {Alzoubi}},\ }\bibfield  {title} {\enquote {\bibinfo {title} {Silibinin attenuates adipose tissue inflammation and reverses obesity and its complications in diet-induced obesity model in mice},}\ }\href@noop {} {\bibfield  {journal} {\bibinfo  {journal} {BMC Pharmacology and Toxicology}\ }\textbf {\bibinfo {volume} {21}},\ \bibinfo {pages} {1--8} (\bibinfo {year} {2020})}\BibitemShut {NoStop}%
\bibitem [{\citenamefont {Ilina}\ \emph {et~al.}(2020)\citenamefont {Ilina}, \citenamefont {Gritsenko}, \citenamefont {Syga}, \citenamefont {Lippoldt}, \citenamefont {Porta}, \citenamefont {Chepizhko}, \citenamefont {Grosser}, \citenamefont {Vullings}, \citenamefont {Bakker}, \citenamefont {Starruß}, \citenamefont {Bult}, \citenamefont {Zapperi}, \citenamefont {Käs}, \citenamefont {Deutsch},\ and\ \citenamefont {Friedl}}]{Ilina2020}%
  \BibitemOpen
  \bibfield  {author} {\bibinfo {author} {\bibfnamefont {O.}~\bibnamefont {Ilina}}, \bibinfo {author} {\bibfnamefont {P.~G.}\ \bibnamefont {Gritsenko}}, \bibinfo {author} {\bibfnamefont {S.}~\bibnamefont {Syga}}, \bibinfo {author} {\bibfnamefont {J.}~\bibnamefont {Lippoldt}}, \bibinfo {author} {\bibfnamefont {C.~A.~L.}\ \bibnamefont {Porta}}, \bibinfo {author} {\bibfnamefont {O.}~\bibnamefont {Chepizhko}}, \bibinfo {author} {\bibfnamefont {S.}~\bibnamefont {Grosser}}, \bibinfo {author} {\bibfnamefont {M.}~\bibnamefont {Vullings}}, \bibinfo {author} {\bibfnamefont {G.~J.}\ \bibnamefont {Bakker}}, \bibinfo {author} {\bibfnamefont {J.}~\bibnamefont {Starruß}}, \bibinfo {author} {\bibfnamefont {P.}~\bibnamefont {Bult}}, \bibinfo {author} {\bibfnamefont {S.}~\bibnamefont {Zapperi}}, \bibinfo {author} {\bibfnamefont {J.~A.}\ \bibnamefont {Käs}}, \bibinfo {author} {\bibfnamefont {A.}~\bibnamefont {Deutsch}}, \ and\ \bibinfo {author} {\bibfnamefont {P.}~\bibnamefont {Friedl}},\ }\bibfield  {title} {\enquote
  {\bibinfo {title} {Cell–cell adhesion and 3d matrix confinement determine jamming transitions in breast cancer invasion},}\ }\href {\doibase 10.1038/s41556-020-0552-6} {\bibfield  {journal} {\bibinfo  {journal} {Nature Cell Biology}\ }\textbf {\bibinfo {volume} {22}},\ \bibinfo {pages} {1103--1115} (\bibinfo {year} {2020})}\BibitemShut {NoStop}%
\bibitem [{\citenamefont {Alkhouli}\ \emph {et~al.}(2013)\citenamefont {Alkhouli}, \citenamefont {Mansfield}, \citenamefont {Green}, \citenamefont {Bell}, \citenamefont {Knight}, \citenamefont {Liversedge}, \citenamefont {Tham}, \citenamefont {Welbourn}, \citenamefont {Shore}, \citenamefont {Kos},\ and\ \citenamefont {Winlove}}]{Alkhouli2013}%
  \BibitemOpen
  \bibfield  {author} {\bibinfo {author} {\bibfnamefont {N.}~\bibnamefont {Alkhouli}}, \bibinfo {author} {\bibfnamefont {J.}~\bibnamefont {Mansfield}}, \bibinfo {author} {\bibfnamefont {E.}~\bibnamefont {Green}}, \bibinfo {author} {\bibfnamefont {J.}~\bibnamefont {Bell}}, \bibinfo {author} {\bibfnamefont {B.}~\bibnamefont {Knight}}, \bibinfo {author} {\bibfnamefont {N.}~\bibnamefont {Liversedge}}, \bibinfo {author} {\bibfnamefont {J.~C.}\ \bibnamefont {Tham}}, \bibinfo {author} {\bibfnamefont {R.}~\bibnamefont {Welbourn}}, \bibinfo {author} {\bibfnamefont {A.~C.}\ \bibnamefont {Shore}}, \bibinfo {author} {\bibfnamefont {K.}~\bibnamefont {Kos}}, \ and\ \bibinfo {author} {\bibfnamefont {C.~P.}\ \bibnamefont {Winlove}},\ }\bibfield  {title} {\enquote {\bibinfo {title} {The mechanical properties of human adipose tissues and their relationships to the structure and composition of the extracellular matrix},}\ }\href@noop {} {\bibfield  {journal} {\bibinfo  {journal} {American Journal of Physiology-Endocrinology and
  Metabolism}\ }\textbf {\bibinfo {volume} {305}},\ \bibinfo {pages} {E1427--E1435} (\bibinfo {year} {2013})}\BibitemShut {NoStop}%
\bibitem [{\citenamefont {Sun}\ \emph {et~al.}(2018)\citenamefont {Sun}, \citenamefont {Liu}, \citenamefont {Wang}, \citenamefont {Deng}, \citenamefont {Zhang}, \citenamefont {Zhao}, \citenamefont {Ma}, \citenamefont {Wu},\ and\ \citenamefont {Sun}}]{sun2018novel}%
  \BibitemOpen
  \bibfield  {author} {\bibinfo {author} {\bibfnamefont {D.}~\bibnamefont {Sun}}, \bibinfo {author} {\bibfnamefont {Y.}~\bibnamefont {Liu}}, \bibinfo {author} {\bibfnamefont {H.}~\bibnamefont {Wang}}, \bibinfo {author} {\bibfnamefont {F.}~\bibnamefont {Deng}}, \bibinfo {author} {\bibfnamefont {Y.}~\bibnamefont {Zhang}}, \bibinfo {author} {\bibfnamefont {S.}~\bibnamefont {Zhao}}, \bibinfo {author} {\bibfnamefont {X.}~\bibnamefont {Ma}}, \bibinfo {author} {\bibfnamefont {H.}~\bibnamefont {Wu}}, \ and\ \bibinfo {author} {\bibfnamefont {G.}~\bibnamefont {Sun}},\ }\bibfield  {title} {\enquote {\bibinfo {title} {Novel decellularized liver matrix-alginate hybrid gel beads for the 3d culture of hepatocellular carcinoma cells},}\ }\href@noop {} {\bibfield  {journal} {\bibinfo  {journal} {International Journal of Biological Macromolecules}\ }\textbf {\bibinfo {volume} {109}},\ \bibinfo {pages} {1154--1163} (\bibinfo {year} {2018})}\BibitemShut {NoStop}%
\bibitem [{\citenamefont {Rahman}\ \emph {et~al.}(2023)\citenamefont {Rahman}, \citenamefont {Wu}, \citenamefont {Chen}, \citenamefont {Sun}, \citenamefont {Liu},\ and\ \citenamefont {Xu}}]{Rahman2023}%
  \BibitemOpen
  \bibfield  {author} {\bibinfo {author} {\bibfnamefont {M.~S.~U.}\ \bibnamefont {Rahman}}, \bibinfo {author} {\bibfnamefont {J.}~\bibnamefont {Wu}}, \bibinfo {author} {\bibfnamefont {H.}~\bibnamefont {Chen}}, \bibinfo {author} {\bibfnamefont {C.}~\bibnamefont {Sun}}, \bibinfo {author} {\bibfnamefont {Y.}~\bibnamefont {Liu}}, \ and\ \bibinfo {author} {\bibfnamefont {S.}~\bibnamefont {Xu}},\ }\bibfield  {title} {\enquote {\bibinfo {title} {Matrix mechanophysical factor: Pore size governs the cell behavior in cancer},}\ }\href {\doibase 10.1080/23746149.2022.2153624} {\bibfield  {journal} {\bibinfo  {journal} {Advances in Physics: X}\ }\textbf {\bibinfo {volume} {8}},\ \bibinfo {pages} {2153624} (\bibinfo {year} {2023})}\BibitemShut {NoStop}%
\bibitem [{\citenamefont {Grosser}\ \emph {et~al.}(2021)\citenamefont {Grosser}, \citenamefont {Lippoldt}, \citenamefont {Oswald}, \citenamefont {Merkel}, \citenamefont {Sussman}, \citenamefont {Renner}, \citenamefont {Gottheil}, \citenamefont {Morawetz}, \citenamefont {Fuhs}, \citenamefont {Xie}, \citenamefont {Pawlizak}, \citenamefont {Fritsch}, \citenamefont {Wolf}, \citenamefont {Horn}, \citenamefont {Briest}, \citenamefont {Aktas}, \citenamefont {Manning},\ and\ \citenamefont {K\"as}}]{grosser2021cell}%
  \BibitemOpen
  \bibfield  {author} {\bibinfo {author} {\bibfnamefont {S.}~\bibnamefont {Grosser}}, \bibinfo {author} {\bibfnamefont {J.}~\bibnamefont {Lippoldt}}, \bibinfo {author} {\bibfnamefont {L.}~\bibnamefont {Oswald}}, \bibinfo {author} {\bibfnamefont {M.}~\bibnamefont {Merkel}}, \bibinfo {author} {\bibfnamefont {D.~M.}\ \bibnamefont {Sussman}}, \bibinfo {author} {\bibfnamefont {F.}~\bibnamefont {Renner}}, \bibinfo {author} {\bibfnamefont {P.}~\bibnamefont {Gottheil}}, \bibinfo {author} {\bibfnamefont {E.~W.}\ \bibnamefont {Morawetz}}, \bibinfo {author} {\bibfnamefont {T.}~\bibnamefont {Fuhs}}, \bibinfo {author} {\bibfnamefont {X.}~\bibnamefont {Xie}}, \bibinfo {author} {\bibfnamefont {S.}~\bibnamefont {Pawlizak}}, \bibinfo {author} {\bibfnamefont {A.~W.}\ \bibnamefont {Fritsch}}, \bibinfo {author} {\bibfnamefont {B.}~\bibnamefont {Wolf}}, \bibinfo {author} {\bibfnamefont {L.-C.}\ \bibnamefont {Horn}}, \bibinfo {author} {\bibfnamefont {S.}~\bibnamefont {Briest}}, \bibinfo {author} {\bibfnamefont {B.}~\bibnamefont
  {Aktas}}, \bibinfo {author} {\bibfnamefont {M.~L.}\ \bibnamefont {Manning}}, \ and\ \bibinfo {author} {\bibfnamefont {J.~A.}\ \bibnamefont {K\"as}},\ }\bibfield  {title} {\enquote {\bibinfo {title} {Cell and nucleus shape as an indicator of tissue fluidity in carcinoma},}\ }\href@noop {} {\bibfield  {journal} {\bibinfo  {journal} {Physical Review X}\ }\textbf {\bibinfo {volume} {11}},\ \bibinfo {pages} {011033} (\bibinfo {year} {2021})}\BibitemShut {NoStop}%
\bibitem [{\citenamefont {Ye}\ and\ \citenamefont {Lin}(2024)}]{Ye2024}%
  \BibitemOpen
  \bibfield  {author} {\bibinfo {author} {\bibfnamefont {Y.}~\bibnamefont {Ye}}\ and\ \bibinfo {author} {\bibfnamefont {J.}~\bibnamefont {Lin}},\ }\bibfield  {title} {\enquote {\bibinfo {title} {Fingering instability accelerates population growth of a proliferating cell collective},}\ }\href {\doibase 10.1103/PhysRevLett.132.018402} {\bibfield  {journal} {\bibinfo  {journal} {Physical Review Letters}\ }\textbf {\bibinfo {volume} {132}},\ \bibinfo {pages} {018402} (\bibinfo {year} {2024})}\BibitemShut {NoStop}%
\end{thebibliography}%
\end{document}